# Physics-Based Decline Curve Analysis and Machine Learning for Temperature Forecasting in Enhanced Geothermal Systems: Utah FORGE

Mina S. Khalaf

University of Houston, Texas, USA

**Abstract**

Reliable temperature forecasting in Enhanced Geothermal Systems (EGS) is essential for reservoir design and economic assessment, yet petroleum-based decline curves and many machine-learning (ML) surrogates do not enforce geothermal heat transfer. In addition, high-fidelity thermo-hydro-mechanical (THM) simulation remains computationally expensive.

This study proposes a unified, physics-consistent framework that advances both decline-curve analysis (DCA) and surrogate modeling for geothermal temperature forecasts. The classical Arps decline family is generalized for geothermal use by introducing an equilibrium-temperature term motivated by Newton-type cooling, ensuring finite late-time temperature limits while reducing exactly to the conventional Arps forms when the equilibrium term is set to zero. The extended decline curves are validated against Utah FORGE downhole temperature measurements and then used to construct learning surrogates on a controlled THM dataset spanning fracture count, well spacing, fracture spacing, host-rock thermal conductivity, and circulation rate. An equation-informed neural network embeds the modified decline equations as differentiable internal computational layers to produce full 0-60 month temperature trajectories from design and operational inputs while preserving interpretable decline structure. A probabilistic Gaussian Process Regression surrogate is also developed for direct multi-horizon forecasting with calibrated uncertainty, while a direct XGBoost regression baseline provides a purely data-driven reference. Across the simulation dataset, the extended decline models reproduce temperature trajectories with near-perfect fidelity (median $R^2$ = 0.999; median RMSE = 0.071 °C), and the equation-informed network achieves typical hold-out errors of MAE = 3.06 °C and RMSE = 4.49 °C. The Gaussian Process surrogate delivers the strongest predictive accuracy across 3-60 month horizons (macro $R^2$ = 0.965; RMSE = 3.39 °C; MAE = 2.34 °C) with well-calibrated uncertainty, whereas the XGBoost baseline exhibits higher errors, underscoring the value of physical structure or probabilistic modeling for reliable geothermal temperature forecasting.

# 1. Introduction

Geothermal energy provides a continuous supply of heat from within the Earth, enabling low-carbon electricity generation and direct heating. Because it operates independently of weather and time of day, it offers reliable baseload power, distinguishing it from wind and solar resources. In recent decades, both research and industry have worked to expand geothermal deployment and extend its use beyond regions with naturally favorable hydrothermal systems. Global assessments show gradual growth, improvements in exploration and drilling technologies, and broader applications for power and heat. However, geothermal still represents only a small share of global electricity generation, indicating considerable untapped potential if technical, economic, and environmental challenges can be overcome [1–4].

Enhanced Geothermal Systems (EGS) extend geothermal use to Hot Dry Rocks (HDR) that lack natural fluid flow. EGS systems improve well performance by extracting heat stored in HDR. These systems rely on drilling and stimulating wells so a closed loop for fluid flow can form. Cool fluid is pumped into the reservoir, moves through the stimulated rock as it warms up, and then returns to the surface. At the surface, heat exchangers capture the thermal energy. Figure 1 shows this EGS setup, including the circulation loop and the surface heat-recovery system. Global reviews of EGS projects and technologies highlight that site conditions matter greatly, especially tectonic setting and stress regime, and that careful stimulation and reservoir management are needed to sustain production. These reviews also chart development strategies and research directions, from exploration databases to cost reduction through improved modeling and maintenance methods [3,5].

Production forecasting in geothermal practice often adapts decline-curve analysis (DCA) from petroleum (exponential, harmonic, hyperbolic), grounded in the conventional Arps equations [6]. However, geothermal behavior can deviate because temperature-controlled heat exchange, injection/production coupling, and fracture-network evolution can shape declines differently than pressure-depletion-dominated oil and gas systems. Case studies show mixed outcomes: near-harmonic trends at The Geysers with spacing and communication effects; alternative, physics-based DCA improving late-time reliability at the Geysers [7,8]; and exponential fits working in some liquid-dominated fields such as Ulubelu [9], underscoring the need to validate any decline form against geothermal data and physics. Recent surrogate and machine-learning (ML) advances for geothermal offer faster and often more accurate forecasts than manual curve fitting, while retaining interpretability when equation structure is embedded. Examples include equation-informed thermal-decline surrogates for optimization, sequence models that emulate hydro-thermal time series, and probabilistic regressors that deliver uncertainty along with predictions [10]. Coupling decline analysis with modern machine learning help decision-makers forecast temperature with reliable uncertainty. Type-curve thinking provides physics-aware diagnostics and guardrails for decline curves, avoiding over-optimistic tails;

machine learning can complement that by learning patterns across scenarios and by surfacing uncertainty. Moreover, fully coupled thermo-hydro-mechanical simulations impose a heavy computational load due to the need to solve nonlinear heat transport, fluid flow, and rock deformation equations simultaneously. This coupling commonly requires multiple hours on a standard workstation [11–13].

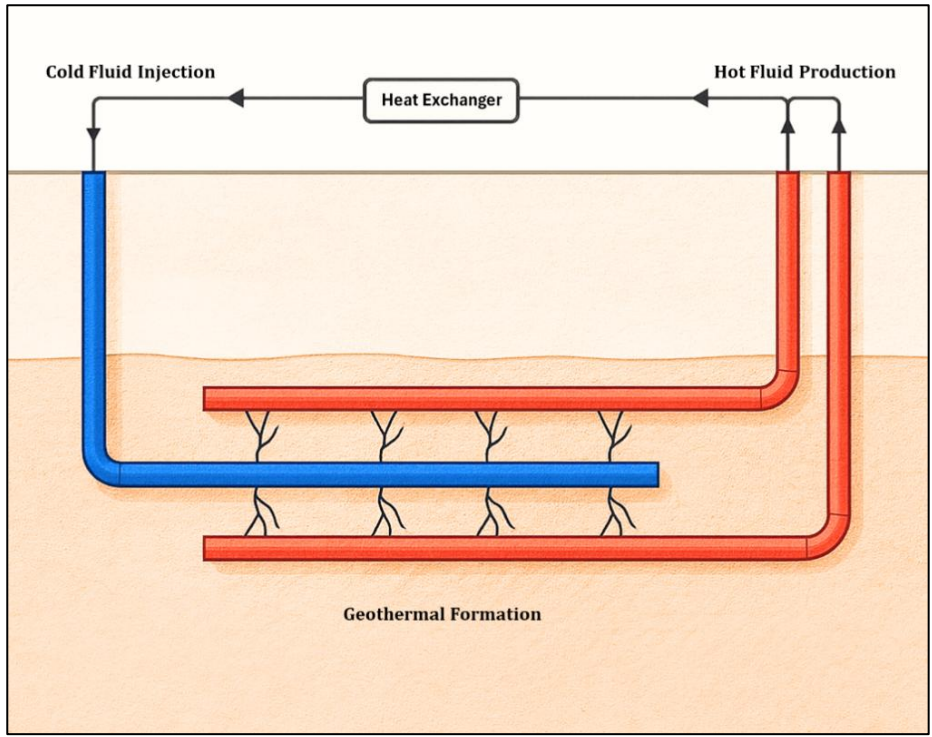

*Figure 1. Schematic of EGS showing horizontal injection and production wells connected through hydraulic fractures.*

This work modifies the Arps equations by introducing an equilibrium-temperature extension based on Newton's cooling law for geothermal applications. The geothermal-extended DCA models are validated using field temperature data from the Utah FORGE project. The study also introduces the use of an equation-embedded neural network for geothermal temperature decline analysis, directly embedding the physical laws into the model for more reliable predictions. Additionally, the study uses a GPR surrogate for fast, uncertainty-aware forecasts. By unifying these approaches on a single thermo-hydro-mechanical (THM) dataset, it offers a coherent and physics-consistent method that reduces the need for computationally heavy simulations.

2. **Literature Review**

Enhanced geothermal systems aim to turn hot but impermeable rock into reliable heat exchangers by creating or improving fluid pathways between wells. Recent reviews describe steady progress from exploration through operation while noting that outcomes depend strongly on stress regime, stimulation strategy, and long-term thermal management [1,3]. studies highlight that stimulation often involves mixed mechanisms: slip on natural fractures together with limited opening and new connections, which helps explain common observations such as flow localization and pressure-limiting behavior [5]. System-level assessments argue that global technical potential is large but constrained by site eligibility and by thermal renewability limits at economic operating conditions, underscoring the need for careful reservoir design and spacing [4]. Advancements in well layouts and staged stimulation suggest practical paths to higher heat extraction in multi-well horizontal configurations, where circulation strategy, fracture properties, and producer–injector distance strongly shape temperature decline [14]. Beyond water, using carbon dioxide as the working fluid is an attractive concept because buoyancy and expansion effects can strengthen natural circulation, reduce parasitic operational load, and raise net heat extraction for comparable pressure drop, though chemical interactions remain a key uncertainty [15].

Conventional decline curve analysis is widely used to summarize production trends and to forecast across oil and gas fields. Arps [6] analyzed exponential, harmonic, and hyperbolic decline models by examining mathematical relationships between production rate, time, cumulative production, and decline percentage. The study also assessed statistical tools like the loss-ratio method, introduced graphical extrapolation techniques, and proposed decline charts to facilitate straight-line predictions [6]. The finding is that decline-curve analysis, particularly the hyperbolic form, provides petroleum engineers with a practical and reliable tool for estimating reserves and future well productivity under capacity production conditions [6]. Fetkovich later showed that these empirical forms sit on a firm physical base by tying them to constant-pressure solutions and log-log type curves, turning decline analysis into a tool that can also diagnose flow mechanisms, estimate permeability, skin, and drainage area, and handle changing backpressure when rates are properly normalized [16,17]. As production moved into tight gas and shale reservoirs, studies on horizontal multi-fractured wells demonstrated that traditional single-segment Arps fits applied during long transient flow can misestimate reserves, and that a diagnostics-first workflow (using tools such as β-derivative plots, normalized-rate plots, and clear identification of flow regimes) should anchor model-based production analysis and then be cross-checked with multiple rate–time decline relations (exponential, hyperbolic, stretched exponential, power-law exponential, Duong, logistic growth, and related forms) to bound time-dependent reserves [18–21]. Field applications of decline curve analysis in oil and gas included Eagle Ford, Bakken, Niobrara, and Middle East tight gas [22–25].

In geothermal settings, case studies show a spectrum of behaviors. At a major vapor-dominated field, normalized analyses and type-curve matches reported trends that sit close to harmonic forms for many wells, with spacing and edge effects shaping individual trajectories [7]. Physics-based refinements have been proposed to improve late-time forecasts by linking rate to the reciprocal of cumulative production, yielding linear trends and interpretable reservoir constants connected to capillary and gravity forces [8]. Applications in liquid-dominated fields show that simple exponential forms can sometimes fit well and support practical planning when declines are smooth and reinjection and make-up drilling are managed coherently, while other geothermal reservoirs are adequately described by exponential or hyperbolic forms for long-term trend work [7,9].

Machine learning and surrogate modeling offer data-efficient ways to emulate coupled thermo-hydraulic behavior and to accelerate design loops. Time-series surrogates that combine recurrent and convolutional layers have been used to predict pressure–temperature evolution, support well-doublet placement, and cut the search cost for economic optima [10]. In addition, recent studies introduced a fast, high-fidelity method to simulate long-term temperature evolution in fractured geothermal reservoirs, integrating complex fracture geometries and injection parameters to support real-time and sustainable operations [26]. ML algorithms have been applied to predict reservoir temperatures from hydrogeochemical data, discussing the viability of ML models and Shapley additive explanation (SHAP) [27,28]. Reduced-basis, physics-respecting surrogates preserve governing structure and deliver orders-of-magnitude speed-ups for multi-query studies such as sensitivity and uncertainty analysis, providing an interpretable alternative to purely black-box networks [12]. Broader reviews chart a surge of geothermal ML across exploration, drilling, reservoir characterization, and production, with sequence models, convolutional networks, and probabilistic learners now common in workflows [29–31]. Related work is also extending ML to operational risk topics such as induced seismicity forecasting, where richer feature sets and attention mechanisms improve short-horizon stability assessment while data scarcity remains a major limitation [32]. Recent work by Yan et al. [33] demonstrated the use of thermal DCA as a surrogate for enhanced geothermal and hot sedimentary aquifer systems by introducing the HyperReLU decline model and training a neural network to predict its parameters from thermo–hydro–mechanical simulation inputs. In their framework, physics enters through the analytical decline structure and an equation-based regularization term in the loss function, while the neural network itself remains a standard fully connected parameter-regression model.

In addition, full-physics simulators are also central to geothermal reservoir engineering because they solve coupled mass, momentum, and energy conservation equations, often with explicit representation of fractures and their mechanical response, so that flow, heat transport, and rock deformation can be predicted in a single framework for realistic reservoir geometries and operating conditions [14,15,34–36]. Although these simulators provide robust predictions of production temperature, pressure, and long-term heat extraction that guide design choices such as circulation rate, fracture aperture and permeability, well spacing and layout, and operating lifetimes, their detailed physics and fine spatial and temporal resolution make them computationally demanding, especially when fractures, complex well layouts, or long production periods are included [11,12,37].

Despite these parallel advances, conventional (petroleum-based) decline-curve equations do not follow geothermal temperature behavior, and most machine-learning surrogates used for geothermal forecasting are not tied to the physics of thermal decline. Because of this, neither approach alone gives reliable, physically consistent temperature forecasts, and there is no unified way to compare or combine them across different EGS conditions. Consequently, it is unclear when conventional decline analysis suffices, when machine learning offers decisive advantages, and whether hybrid approaches can combine the strengths of both.

This study fills the persistent gap between physics-based decline analysis, data-driven surrogates, and high-fidelity thermo-poroelastic simulation by unifying all three within a single forecasting framework. It introduces the equilibrium-temperature extension of the full Arps family, enforcing geothermal heat-transfer behavior while retaining exact reduction to the petroleum forms. It also uses an equation-embedded neural network that directly incorporates these physical laws for reliable predictions. Additionally, the approach includes a GPR surrogate for rapid, uncertainty-aware forecasts. Within a single thermo-hydro-mechanical dataset, it evaluates key geological, design, and operational variables. By combining all these methods, it offers a unified, physics-consistent framework that reduces the need for intensive full-physics simulations.

## 3. EGS System Configuration and Dataset

The dataset used in this study is constructed from 110 high-fidelity simulations representing EGS configurations under varying geological and operational conditions. The EGS simulator used for data generation integrates a fully coupled thermo-hydro-mechanical (THM) framework that simulates mechanical deformation, fluid flow, and heat transfer in geothermal reservoirs using the Displacement Discontinuity Method (DDM) formulation. The model includes within-fracture flow governed by Poiseuille's law. Thermal transport within-fracture accounts for conduction, convection, and rock-fluid interactions [13]. The primary objective in generating this dataset was to capture a broad range of temperature decline curves resulting from the interaction between fracture geometry, rock thermal properties, and fluid injection parameters. By systematically varying these factors, the dataset provides a comprehensive foundation for evaluating and comparing both conventional decline curve models and modern machine learning surrogates in forecasting reservoir temperature evolution.

This study uses a doublet EGS configuration with one injection well and one production well. The wells are connected through hydraulic fractures that provide the flow path for fluid circulation between them, and the produced hot fluid is routed to the surface for heat recovery. The formation temperature ($T_{res}$) is 146 °C and the injection fluid temperature is 40 °C. The simulations are based on single-phase fluid flow, using water as the working fluid. Under reservoir conditions of 146 °C and 10 MPa, the injected water ($T_{inj}$) at 40 °C remains in liquid form. The geothermal formation properties are listed in Table 1.

*Table 1. Rock properties of Westerly Granite* [38].

| $\nu$ | $\nu_u$ | $G$, GPa | $k$, md | $\phi$, % | $\mu$, cp | $c_f$, psi$^{-1}$ | $B$ | $T_{res}$, °C |
|---|---|---|---|---|---|---|---|---|
| 0.25 | 0.33 | 15 | 0.0005 | 1 | 0.3 | $2E-6$ | 0.85 | 146 |
| $\sigma_y$, MPa | $\sigma_x$, MPa | $\sigma_{xy}$, MPa | $p_r$, MPa | $\alpha$ | $B_f, \frac{1}{°C}$ | $B_s, \frac{1}{°C}$ | $C_T, \frac{m^2}{s}$ | $T_{inj}$, °C |
| 20 | 13 | 0 | 10 | 0.44 | $2.4E-4$ | $2.1E-5$ | $1.1E-6$ | 40 |

Each simulation case corresponds to a unique EGS configuration defined by five design variables: the number of hydraulic fractures, well spacing, fracture spacing, host-rock thermal conductivity, and fluid circulation rate. These variables were selected because they represent the dominant physical and operational parameters influencing thermal performance and long-term heat extraction in fractured geothermal reservoirs. The number of hydraulic fractures was set to one, two, or three to capture different levels of complexity in flow pathways (Figure 2). Well spacing ranged from 50 to 250 meters, reflecting field-scale separations between injection and production wells (Figure 3). Fracture spacing was varied between 10 and 100 meters in multi-fracture configurations, while single-fracture cases were assigned a spacing value of zero (Figure 4). Fluid circulation rates varied between 30 and 300 barrels per day per meter-thickness (Figure 5). Thermal conductivity values were sampled from 1.5 to 15 W/(m·K), encompassing typical ranges observed in crystalline basement formations and ensuring that both low- and high-conductivity rock types were represented (Figure 6).

To facilitate structured analysis, the dataset was categorized according to the number of hydraulic fractures (Figure 2). In the single-fracture group, well spacing was relatively large, with an average of 143 meters, while thermal conductivity averaged 8.1 W/(m·K) and fluid circulation rates were comparatively high, averaging 152.4 barrels per day per meter (Table 2). The absence of fracture spacing in this configuration simplified the heat exchange geometry, emphasizing the effects of well spacing and thermal conductivity. In the two-fracture group, the mean well spacing decreased to approximately 104 meters, and fracture spacing was introduced with an average of 47 meters. The mean thermal conductivity in this group was slightly higher at 9.2 W/m·K, while fluid circulation rates averaged 144 barrels per day per meter. The three-fracture group further reduced well spacing to an average of 86 meters, while maintaining similar fracture spacing (mean 46 meters) and thermal conductivity (mean 8.6 W/m·K). Fluid circulation rates remained comparable at an average of 143 barrels per day per meter, suggesting that once multiple fractures were present, the number of fractures had a weaker influence on required injection magnitude.

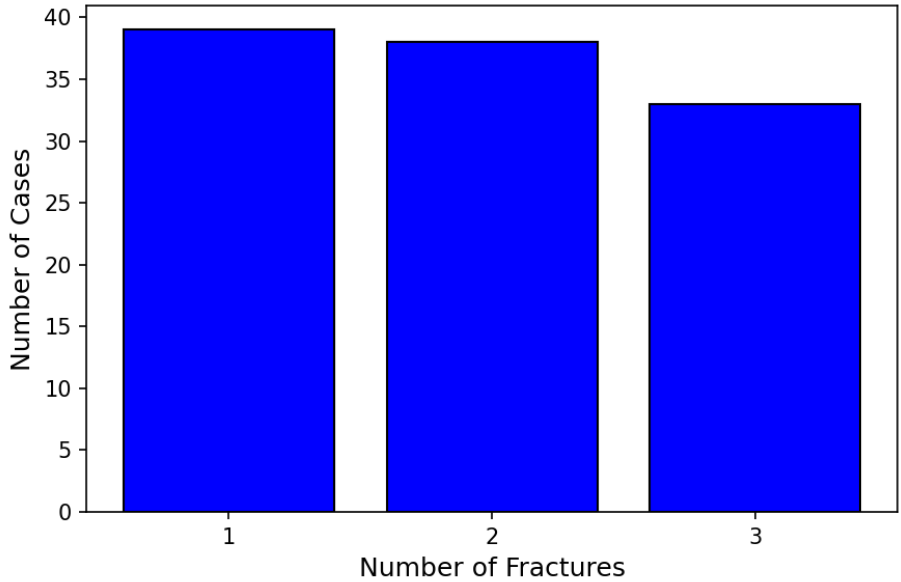

*Figure 2. Number of cases per* hydraulic *fracture count.*

*Table 2. Summary statistics by fracture count for well spacing, fracture spacing, thermal conductivity, and fluid circulation rate.*

|  |  | mean | std | min | q25 | q50 | q75 | max |
|---|---|---|---|---|---|---|---|---|
| **Single fracture** | Well spacing, m | 143.1 | 58.5 | 50 | 100 | 130 | 200 | 250 |
|  | Thermal conductivity, W/(m.K) | 8.1 | 4.1 | 1.5 | 4.8 | 8.4 | 10.7 | 15 |
|  | Fluid circulation rate, bbl/day/m | 152.4 | 74.6 | 37.5 | 100 | 147.4 | 209.3 | 299.4 |
| **Two fractures** | Well spacing, m | 103.9 | 32.3 | 50 | 82.5 | 100 | 120 | 200 |
|  | Fracture spacing, m | 47.3 | 25.3 | 10 | 30 | 39.2 | 64.4 | 99.8 |
|  | Thermal conductivity, W/(m.K) | 9.2 | 3.9 | 1.8 | 6.6 | 10.7 | 10.9 | 15 |
|  | Fluid circulation rate, bbl/day/m | 143.9 | 72.2 | 30.8 | 100 | 116.9 | 193.9 | 299.4 |
| **Three fractures** | Well spacing, m | 85.8 | 29.4 | 50 | 70 | 80 | 100 | 200 |
|  | Fracture spacing, m | 45.5 | 25.4 | 10 | 30 | 30 | 64.9 | 98.9 |
|  | Thermal conductivity, W/(m.K) | 8.6 | 4 | 1.5 | 5.1 | 10.7 | 10.7 | 15 |
|  | Fluid circulation rate, bbl/day/m | 142.7 | 74.3 | 35 | 100 | 117.7 | 200 | 296.8 |

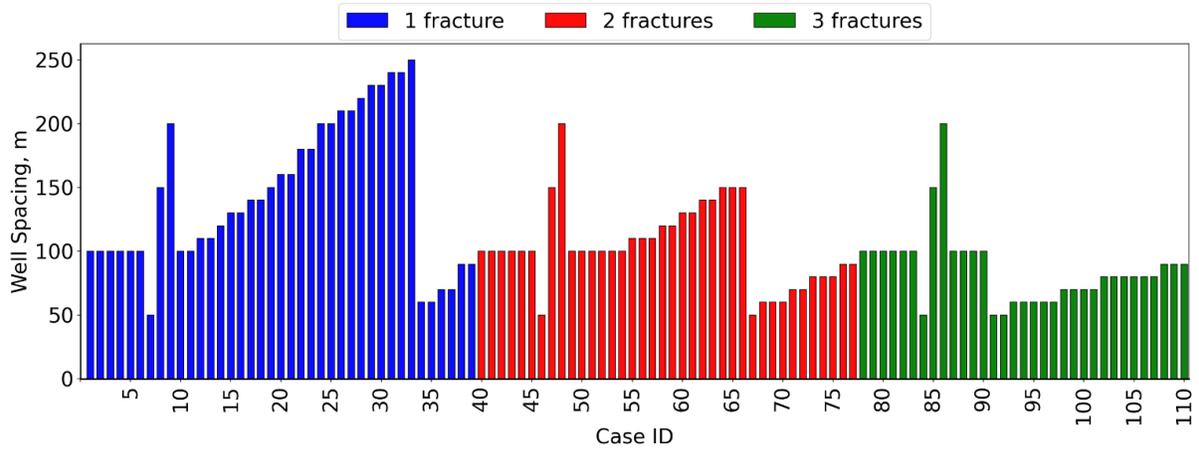

*Figure 3.* Distribution of well spacing in meters across the 110 cases.

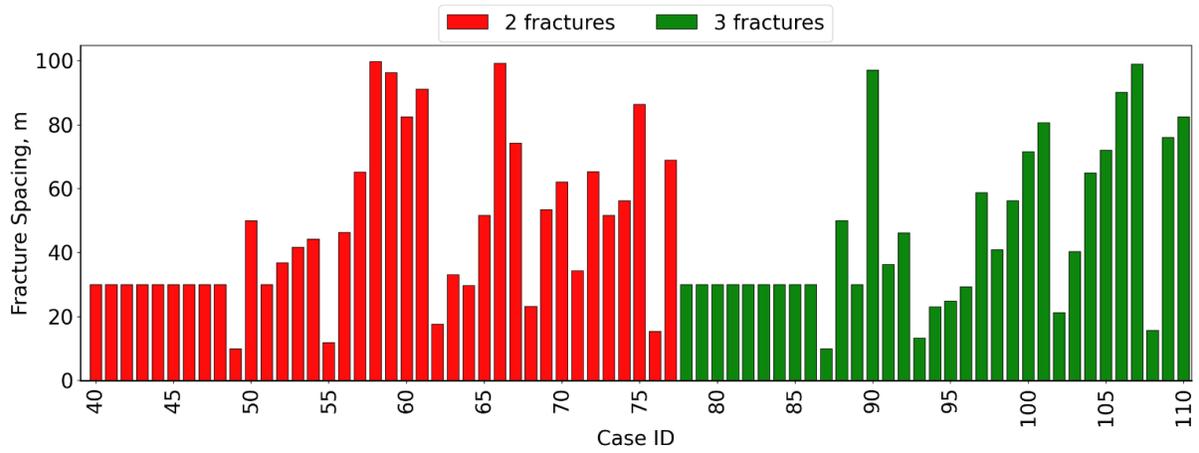

*Figure 4.* Distribution of fracture spacing in meters for 2- and 3-fracture cases.

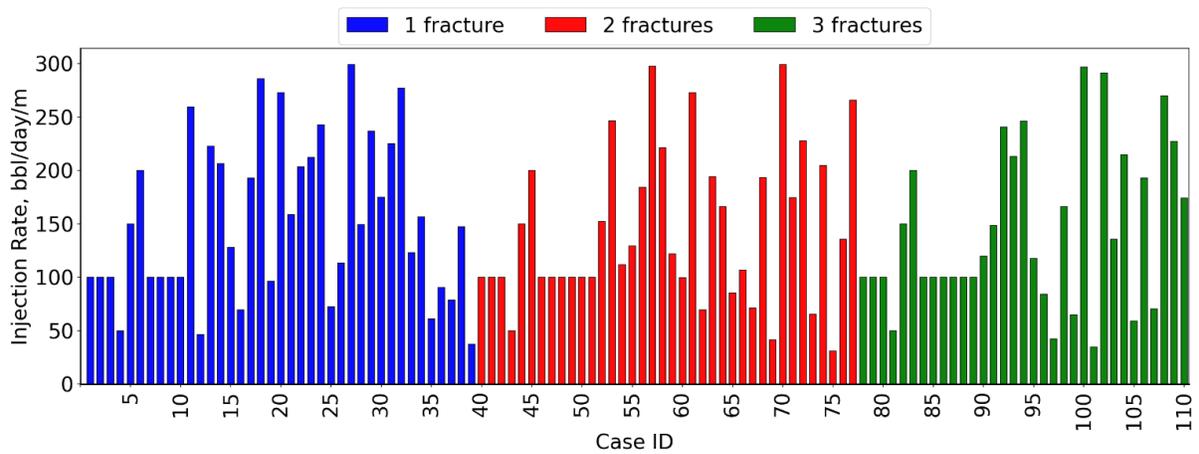

*Figure 5.* Distribution of fluid circulation rate in bbl/day/m-thickness for 110 cases.

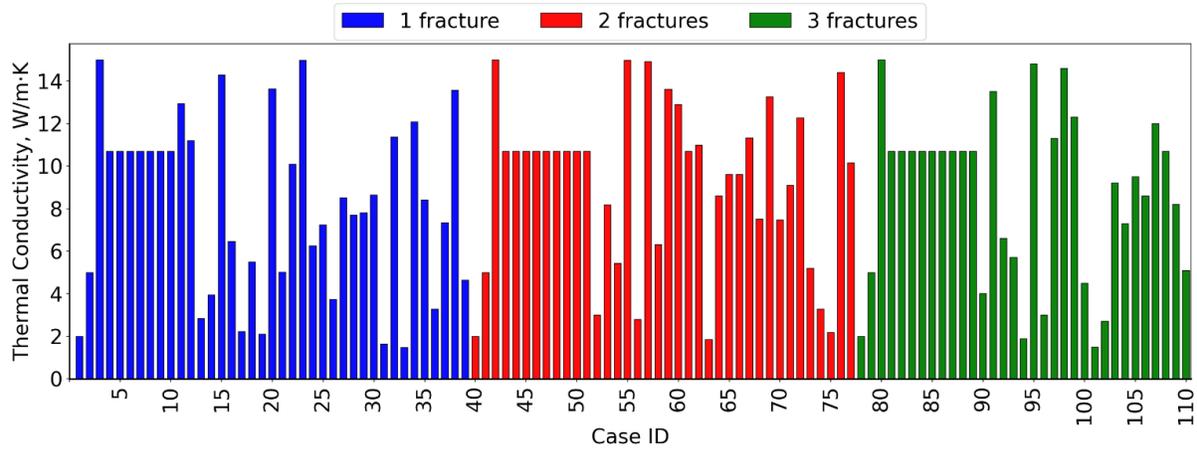

*Figure 6. Distribution of rock thermal conductivity for the 110 cases.*

Each simulated case produces a temperature-time profile capturing the evolution of reservoir temperature (Figure 7). The time series are selected at discrete monitoring intervals (of 3-month periods) from 3 to 60 months, providing the dependent variables for both analytical and machine learning models. By combining a controlled design matrix of geological and operational variables with consistent temporal outputs, the dataset supports robust statistical comparison of forecasting performance across diverse decline mechanisms and modeling frameworks. The resulting collection represents a comprehensive benchmark for testing the interpretability, accuracy, and generalization of temperature decline models in engineered geothermal reservoirs.

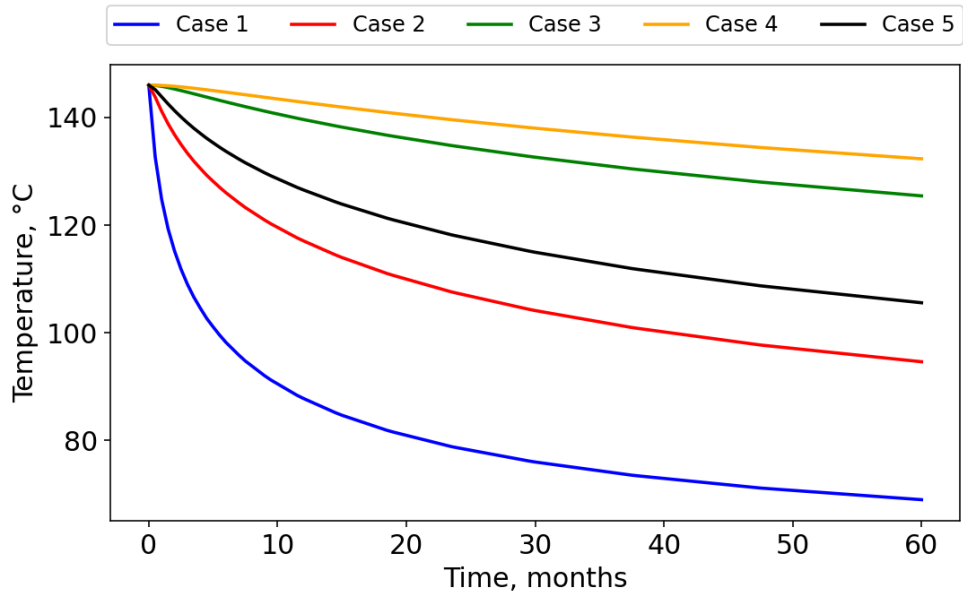

*Figure 7. Temperature versus time for five different cases.*

As shown in the correlation matrix (Figure 8), the production temperature is strongly and consistently tied to rock thermal conductivity (0.56-0.61 across 3-60 months), followed by a growing positive association with well spacing (0.31 at 3 months rising to ≈0.42 by 60 months). Fracture spacing shows a small positive effect that slightly tapers over time (from 0.12 to 0.10). Fluid circulation rate is moderately negative and strengthens with time (-0.39 to -0.44), suggesting higher rates coincide with lower temperatures in this dataset. The number of fractures is near-neutral and drifts from a tiny positive to a slight negative (0.05 to -0.02). Overall, the temperature signal is dominated by thermal conductivity and well spacing (positive) and countered by fluid rate (negative), while fracture spacing and fracture count play secondary roles.

*Figure 8. Correlation matrix of the dataset variables.*

## 4. Methods

This section describes the modeling steps used to analyze and forecast temperature decline. The workflow has four parts. First, the modified decline-curve equations are applied to each temperature-time series to obtain compact, physics-consistent summaries of thermal behavior. Second, an equation-informed neural network is trained to map reservoir and operating inputs to full temperature trajectories while keeping the analytical structure of the DCA forms. Third, a Gaussian Process Regression model is built to give direct multi-horizon forecasts with uncertainty. Finally, an XGBoost regression model serves as a direct, data-driven multi-horizon baseline These methods allow a clear comparison between physics-based fits, physics-guided learning, flexible statistical prediction, and purely data-driven models.

## 4.1. Geothermal-Extended Decline Curve Models

This study extends the Arps-type decline equations (petroleum-based) [6,21,24,39,40] by introducing an equilibrium temperature term ($T_\infty$) for applications to geothermal systems. This modification is applied to all decline forms (exponential, harmonic, hyperbolic, and stretched exponential) to represent the surrounding rock thermal baseline. The inclusion of this term allows the temperature to approach a finite equilibrium value rather than zero, accurately reflecting the physical behavior of geothermal systems. If the equilibrium temperature is set to zero, the equations simplify directly to the original Arps-type DCA forms commonly used in oil and gas production, where the variable of interest declines asymptotically to zero. The formulation follows Newton's Law of Cooling [41,42], which describes how temperature gradually approaches its ambient or equilibrium value over time, ensuring physical consistency with geothermal heat-transfer processes. This modification was introduced to overcome the limitation of the DCA forms in representing the early temperatures before thermal breakthrough in geothermal systems.

The modified decline-curve analysis was applied to characterize the temporal behavior of production temperature using these four modified models: exponential, harmonic, hyperbolic, and stretched exponential. These models describe how temperature evolves with time under different physical assumptions, providing a quantitative framework for decline trend interpretation. Parameters for each model were determined through nonlinear least-squares optimization under bounded constraints to ensure realistic and stable solutions. Model performance was evaluated primarily using the Akaike Information Criterion (AIC) [43], which balances fit quality with parameter simplicity, while additional metrics, including $R^2$ and RMSE, were included for diagnostic interpretation.

The exponential model is a three-parameter model that assumes a constant fractional rate of decline over time, making it one of the simplest and most widely used DCA formulations. It describes systems in which the rate of temperature change is proportional to the remaining temperature difference between the initial and equilibrium states. This model is suitable for processes dominated by steady, continuous decay mechanisms such as conductive cooling or exponential depletion. The modified exponential model is given as:

$$T(t) \;=\; T_\infty + (T^0 - T_\infty)\, e^{-D\, t} \qquad (1)$$

In this model, T(t) represents the temperature at time t; $T_0$ is the initial temperature at t = 0; $T_\infty$ is the asymptotic or equilibrium temperature that the system approaches at large times; D is the exponential decline rate constant that governs the speed of temperature decay; and t denotes time.

The harmonic model is a three-parameter model that extends the exponential formulation by allowing a slower, time-dependent rate of decline. It assumes that the rate of temperature change decreases inversely with time, capturing cases where decline decelerates gradually rather than following a constant exponential trend. This behavior often represents systems with diminishing driving forces or increasing resistance to heat transfer. The modified harmonic model can be expressed as:

$$T(t) = T_\infty + \frac{T^0 - T_\infty}{1 + D t} \tag{2}$$

Here, T(t) is the temperature as a function of time t; $T_0$ is the initial temperature; $T_\infty$ is the long-term asymptotic temperature; and D is the harmonic decline constant that determines how quickly the temperature decreases with time. This model produces a slower decline rate compared to the exponential model, reflecting a gradual reduction over time. The harmonic model includes the same temperature parameters as the exponential case, with a decline constant that controls the curvature of the decline.

The hyperbolic model is a four-parameter model that introduces an additional curvature parameter that generalizes both exponential and harmonic forms. This flexibility enables the model to describe a wider range of decline curves, from rapidly decaying to gently tapering trends. The hyperbolic exponent determines how sharply the decline rate changes with time. It is frequently used in subsurface engineering and reservoir studies to capture complex, nonlinear decay patterns. The modified hyperbolic model can be written as:

$$T(t) = T_\infty + (T^0 - T_\infty)(1 + b D t)^{-\frac{1}{b}} \tag{3}$$

In this model, T(t) is the temperature at time t; $T_0$ represents the initial temperature; $T_\infty$ is the final or equilibrium temperature; D is the decline constant; b is the hyperbolic exponent that controls the curvature of the decline curve; and t is time. When $b \to 0$, the model approaches the exponential form, and when b = 1, it becomes harmonic, providing flexibility to represent intermediate decline curves. The model parameters define the initial and asymptotic temperatures, the nominal decline constant, and a shape factor that governs the curvature transition.

The stretched exponential model is a four-parameter model that represents a more generalized form of exponential decay that accounts for a distribution of decline rates rather than a single characteristic rate. This model is particularly effective for systems exhibiting nonuniform relaxation or diffusive processes, where early-time and late-time behaviors differ significantly. The stretched exponential model can be written as:

$$T(t) = T_\infty + (T^0 - T_\infty) \exp(-(D t)^\beta) \tag{4}$$

Here, T(t) is the temperature as a function of time t; $T_0$ is the initial temperature; $T_\infty$ is the asymptotic temperature; D is a rate constant that defines the characteristic timescale of decline; β is the stretching exponent controlling the curvature and deviation from simple exponential behavior; and t is time. This model captures systems where decline rates vary with time, allowing slower, nonuniform decay. Its parameters include the initial and asymptotic temperatures, a rate constant, and a stretching exponent that defines the degree of deviation from pure exponential behavior.

All models were fitted by minimizing the residual sum of squares (RSS) between observed and predicted temperatures, ensuring the best possible fit within the defined bounds.

$$RSS = \Sigma (y_i - \hat{y}_i)^2 \qquad (5)$$

In this expression, $y_i$ are the observed temperature data points, and $\hat{y}_i$ are the corresponding model-predicted values at the same times $t_i$. The RSS quantifies the total deviation between observations and model predictions, serving as the objective function minimized during curve fitting.

Model performance was compared using the Akaike Information Criterion [43], which rewards goodness of fit while penalizing excessive model complexity.

$$AIC = n\, ln\left(\frac{RSS}{n}\right) + 2k \qquad (6)$$

In this equation, n is the number of data points, RSS is the residual sum of squares, and k is the number of fitted parameters. AIC measures the relative quality of a model, balancing goodness of fit and complexity. Lower AIC values indicated models that achieved an optimal trade-off between fit quality and simplicity, guiding the selection of the most representative decline model for each case.

### 4.2. Equation-Informed Neural Network (EINN)

The equation-informed model is implemented as a neural network (NN) applied to the temperature-decline outputs generated by the modified DCA formulations (Section 4.1). This allows the NN to learn from physically based decline patterns while preserving the underlying trends and parameters defined by DCA. This equation-informed supervised model was developed to predict decline-curve parameters and the corresponding temperature trajectories on a fixed time grid. Inputs comprised five predictors (number of fractures, well spacing, fracture spacing, thermal conductivity, and fluid circulation rate). Targets followed the four canonical DCA equations (exponential, harmonic, hyperbolic, stretched-exponential), each parameterized by its standard symbols (e.g., T∞, D, b, β). All feature preprocessing, including log-transformation and standardization, was fit exclusively on the training subset within each split or cross-validation fold and then applied unchanged to validation and test subsets, preventing information leakage from held-out cases. The temporal domain spanned 0–60 months in 0.5-month increments with a fixed initial temperature $T^0$ = 146.0.

To prevent scenario-level data leakage, all data partitions for the equation-informed neural network were performed at the thermo–hydro–mechanical simulation case level. Each THM case generates multiple samples corresponding to different decline-curve formulations; therefore, all samples derived from the same simulation case were kept entirely within a single partition. An independent group-wise hold-out split was first applied, reserving 20% of simulation cases as a test set and assigning the remaining 80% of cases to training and validation. Within the training-validation pool, group K-fold cross-validation (K = 5) was used, ensuring that no simulation case appeared in both training and validation folds.

The predictor network is a feed-forward, multi-head neural network consisting of a shared trunk with two fully connected layers (each of 128 neurons with ReLU activations) and distinct single-layer linear output heads, each corresponding to one DCA model (exponential, harmonic, hyperbolic, and stretched-exponential). Each head predicts the set of parameters specific to its decline-curve equation: two parameters for the exponential and harmonic forms, and three for the hyperbolic, and stretched-exponential forms. The initial reservoir temperature, $T^0$, is not predicted in this ML model, as it is a known value representing the starting thermal condition of the reservoir. The predicted parameters are passed through bounded transformations (using softplus and scaled-sigmoid) to ensure physical validity, such as positivity of decline rates and asymptotes and bounded b and β. Given the parameters of one head, an equation-informed computational layer computes the complete temperature-time trajectory on the 0–60-month grid using the corresponding DCA equation.

Training optimized a composite objective combining a time-weighted curve discrepancy and a parameter consistency term. The curve term was a Huber loss (δ = 1.0) applied pointwise on the time grid and multiplied by normalized temporal weights w(t) = 1 / (1 + t/τ) with τ = 18 months, averaged over time. The parameter term was a mean-squared error between the head predicted parameters and the reference DCA parameters. The total loss per sample was:

$$L = \underbrace{Huber_\delta(T_{pred}(t), T_{ref}(t))\ averaged\ with\ w(t)}_{curve\ loss} + 0.2 \times \underbrace{||\theta_{pred} - \theta_{ref}||^2}_{parameter\ loss} \qquad (7)$$

In this equation, $T_{pred}$ and $T_{ref}$ denote the predicted and reference temperature trajectories as functions of time t; $Huber_\delta$ is the Huber loss function with transition parameter δ=1.0, combining quadratic and linear regimes for residuals of different magnitudes; $w(t) = 1/(1 + (t/\tau))$ is a temporal weighting function with τ= 18 months used to emphasize early-time accuracy; $\theta_{pred}$ and $\theta_{ref}$ represent vectors of the predicted and reference DCA parameters, respectively; $||\theta_{pred} - \theta_{ref}||^2$ denotes the mean-squared difference between parameter vectors; and the coefficient 0.2 scales the parameter-consistency term relative to the curve-fitting term to maintain balance between temperature-curve fidelity and parameter accuracy.

Mini-batches were balanced across model equations by simple class-wise oversampling. Optimization used Adam with learning rate $10^{-3}$, weight decay $10^{-4}$, batch size 64, and 600 epochs. A Reduce-on-Plateau scheduler (factor 0.5, patience 15) adjusted the learning rate based on validation loss.

Evaluation compared the rendered ML curves to the reference DCA curves using RMSE and MAE. After cross-validation, a final model was trained on the full train+validation set and assessed on the 20% test partition with the same metrics, reported per equation and overall (mean ± SD). Outputs included parameter predictions for all cases, per-case curve plots (reference vs. ML), and scatter plots of true vs. predicted temperatures at 15, 30, 45, and 60 months.

### 4.3. Gaussian Process Regression (GPR)

Gaussian Process Regression (GPR) was applied independently to each prediction horizon (3-month increments from 3 to 60 months) to model temperature evolution as a function of five predictors: number of fractures, well spacing, fracture spacing, thermal conductivity, and fluid circulation rate. The GPR dataset consists of one observation per simulation case, where each case is characterized by a single set of input variables and a vector of temperature targets at multiple time horizons. An independent hold-out split was performed by case identity, with 80% of cases used for training and 20% reserved as an unseen test set (Figure 9). This case-wise evaluation strategy prevents information leakage across correlated temperature horizons and ensures that reported GPR performance reflects generalization to unseen geothermal system configurations.

During model development, K-fold cross-validation was performed within the training cases only to estimate generalization performance and produce out-of-fold (OOF) predictions. In each fold, the feature set was standardized using a z-score transformation derived solely from the fold training data, such that each feature has a mean of 0 and a standard deviation of 1. After cross-validation, the scaler was refitted on the complete training data prior to final model testing.

Each target variable was modeled using a Gaussian Process Regressor with a composite covariance kernel consisting of a constant kernel multiplied by a radial basis function (RBF) kernel, combined additively with a linear (dot product) component and a white-noise term. This structure allows the model to capture both nonlinear smooth variations and linear global trends. The model used normalized outputs and a numerical regularization parameter $\alpha = 10^{-6}$.

The predictive performance was evaluated using the coefficient of determination ($R^2$), root mean squared error (RMSE), and mean absolute error (MAE). Metrics were computed for each fold and aggregated as mean and standard deviation across folds to summarize training performance. After training on the full training dataset, model accuracy was assessed on the independent test partition using the same metrics. Predictive uncertainty was derived from the Gaussian process posterior variance, and 95% prediction intervals were calculated as $\mu \pm 1.96\, \sigma$, where $\mu$ is the mean and $\sigma$ is the standard deviation. Diagnostic outputs include scatter plots comparing predicted and true temperature values at selected horizons (15, 30, 45, and 60 months).

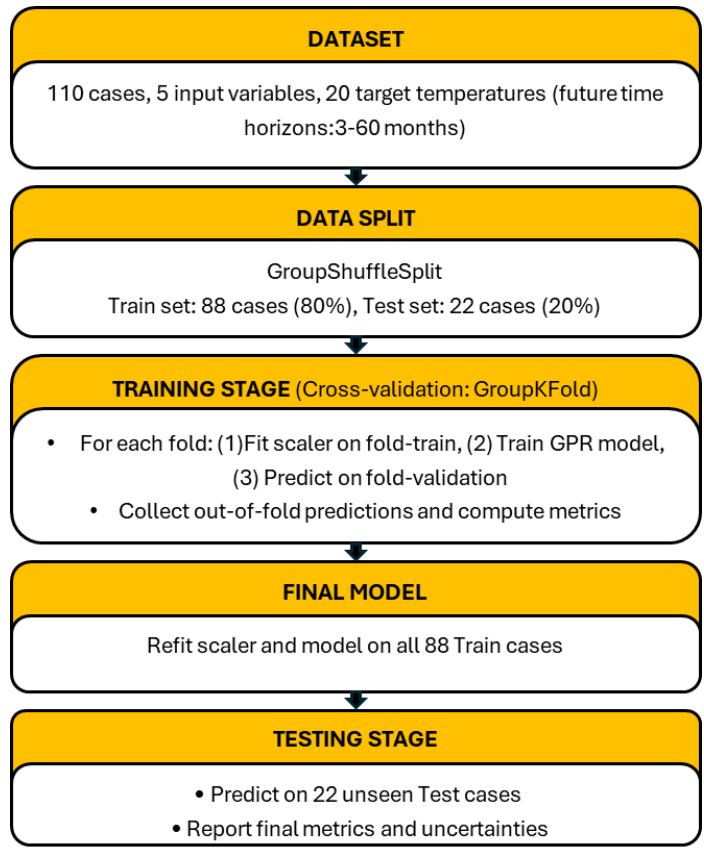

*Figure 9.* GPR schematic showing data structure, data split into training and hold-out sets, and cross-validation within training to produce OOF predictions.

### 4.4. Direct XGBoost

A deterministic gradient-boosted decision tree model based on XGBoost is used as a non-physics-informed baseline for temperature forecasting. The implementation is designed to mirror the data handling, splitting strategy, and evaluation protocol of the other surrogate models, ensuring a fair and consistent comparison. The prediction task is formulated as direct multi-horizon regression. Reservoir temperatures at fixed times, ranging from 3 to 60 months at 3-month intervals, are predicted independently. Each forecast horizon is treated as a separate regression problem using the same set of input features (number of hydraulic fractures, fracture length, fracture spacing, host-rock thermal conductivity, and injection rate), which describe fracture characteristics, thermal properties, and operational conditions. Each case corresponds to a single THM simulation and produces one temperature value per forecast horizon.

To prevent information leakage across correlated samples, all data splitting is performed at the simulation-case level using a group identifier. An unseen hold-out test set comprising 20% of the cases is first constructed using a group-aware shuffle split. The remaining 80% of cases form the training pool. Within the training pool, model evaluation is carried out using group-aware K-fold cross-validation, ensuring that all samples from a given simulation case appear exclusively in either the training or validation fold. This approach preserves statistical independence between folds while allowing robust estimation of generalization performance.

XGBoost regressors are trained with a squared-error objective and histogram-based tree construction for computational efficiency. Two predefined configurations are supported: a moderate-capacity configuration for faster training and a higher-capacity configuration with smaller learning rate and deeper trees. Random seeds are fixed to ensure reproducibility across runs. Overfitting is controlled through early stopping applied in a group-aware manner. Within each training fold, an inner group-based validation split is created from the training portion only. Model training terminates when validation RMSE fails to improve for a specified number of boosting rounds. A compatibility wrapper is used to ensure that early stopping functions correctly across different XGBoost versions. This procedure is applied both during cross-validation and during final training on the full training set prior to testing.

Model performance is evaluated in two stages. First, out-of-fold predictions from GroupKFold cross-validation are used to compute unbiased training metrics. Second, models retrained on the full training set are evaluated once on the unseen hold-out test set. Performance is quantified using the coefficient of determination ($R^2$), root mean squared error (RMSE), and mean absolute error (MAE), reported per forecast horizon and summarized using macro-averaged metrics across all horizons.

## 5. Validation of Geothermal-Extended DCA Models

This section presents the validation of the geothermal-extended decline-curve analysis models using a field case from the Utah FORGE project. The validation dataset consists of downhole pressure-temperature measurements acquired via fiber-optic gauges installed in Utah FORGE Well 16B(78)-32 at a measured depth of 7,056.67 feet [44]. The data comprise time-series pressure and temperature records collected in April 2024 (Figure 10). Well 16B(78)-32 functions as the production well for reservoir creation, fluid circulation, and heat-extraction demonstration at the FORGE site. The well was drilled as part of a doublet configuration, located approximately 300 feet above and parallel to the injection well 16A(78)-32. Although the planned total depth was 10,658 feet, drilling exceeded this target; operations began on April 26th, 2023, and by June 20th, 2023, the measured depth reached 10,947 feet with a vertical depth of 8,357 feet. Drilling operations included a field trial of insulated drill pipe (IDP) supplied by Eavor Technologies [45]. The IDP limits counter-current heat transfer between cooler drilling fluid within the drill string and hotter annular returns, ensuring that the bottomhole assembly remains immersed in relatively cool fluid. A total of 350 joints of IDP were deployed across two successive bottomhole assemblies during drilling [45].

Figure 10 presents downhole temperature on the left vertical axis (°F) and pressure on the right vertical axis (psi) as functions of local date and time. The records capture operational transients over the entire flow period, including intervals where production pressure remains relatively stable. Production pressure changes are accompanied by temperature responses, whereas other segments show more gradual thermal evolution under nearly constant pressure. These patterns indicate shifts in flow and operating conditions during well activity and form the empirical foundation for the decline-curve fitting and validation performed in the following analysis.

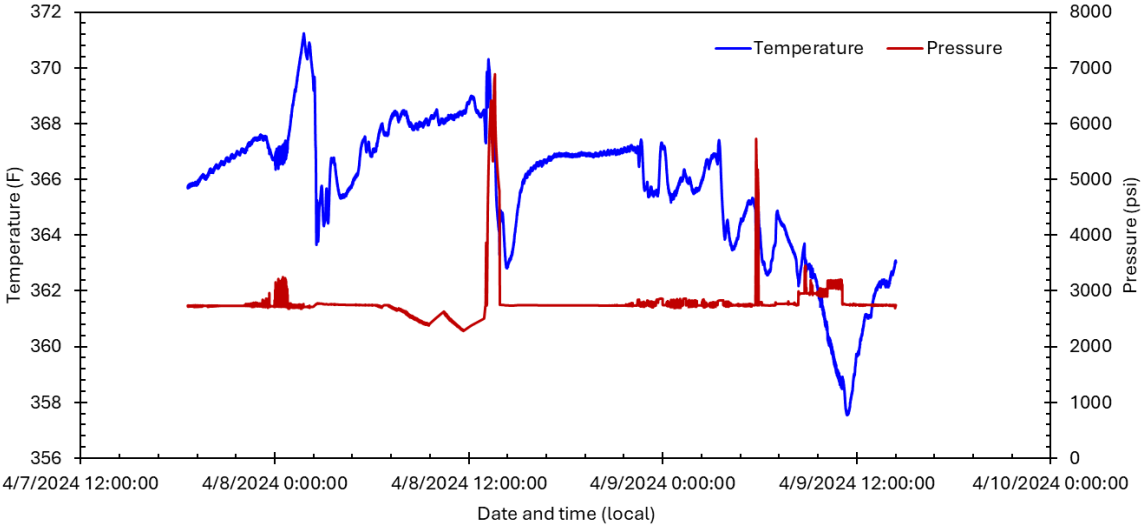

*Figure 10. Downhole pressure-temperature field measurements of Utah FORGE Well 16B(78)-32.*

Figure 11 presents a comparison between the measured downhole temperature data and the corresponding fits obtained using both the conventional (petroleum-based) and the modified decline-curve (geothermal-extended) formulations. The temperature observations are displayed as blue data points, whereas the conventional and geothermal-extended models are shown as continuous curves. Each subplot represents a different decline-curve form, including the exponential, harmonic, hyperbolic, and stretched exponential models. In all panels, the same temperature time series data is used for model fitting, enabling a direct and consistent comparison of how effectively each formulation captures the overall temperature-decline behavior. The divergence between the conventional and modified formulations becomes increasingly evident at later times, where differences in curvature and decline rate lead to noticeably different representations of the observed temperature trend.

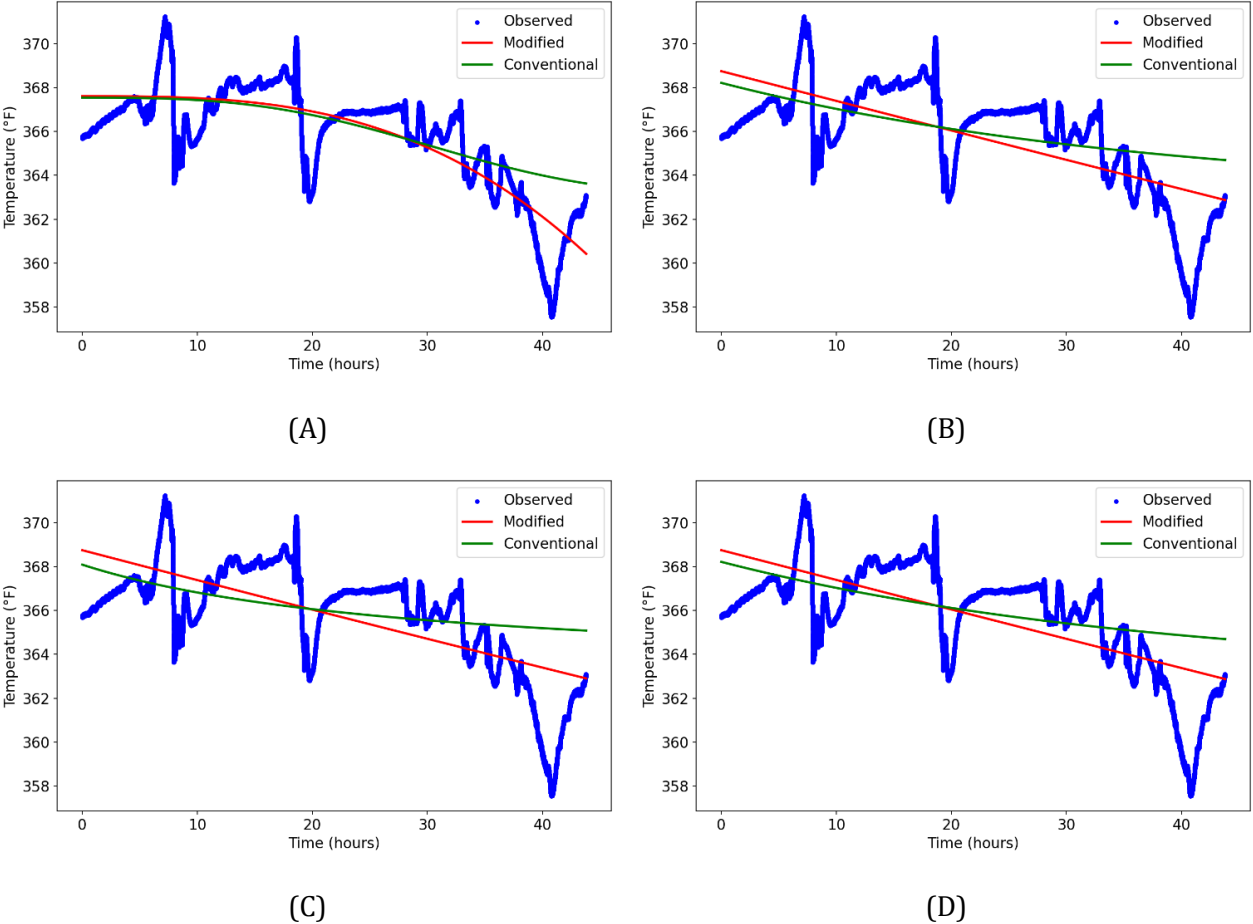

(A)  (B)

(C)  (D)

*Figure 11. Temperature data with conventional and modified decline-curve models: (A) Stretched exponential, (B) Hyperbolic, (C) Harmonic, and (D) Exponential.*

Figure 12 summarizes the statistical performance of the conventional and modified decline-curve models applied to the temperature data shown in Figure 11, using the coefficient of determination ($R^2$), root mean square error (RMSE), and Akaike Information Criterion (AIC) as quantitative measures of fit quality and model parsimony. Across all decline formulations (exponential, harmonic, hyperbolic, and stretched exponential) the modified models consistently outperform their conventional counterparts, exhibiting higher $R^2$ values together with lower RMSE and AIC values. For example, the modified exponential model increases $R^2$ from approximately 0.31 to 0.47 while reducing RMSE from about 2.06 to 1.81, accompanied by a substantial decrease in AIC. Similar improvements are observed for the harmonic and hyperbolic formulations, where gains in explained variance and reductions in error are matched by lower AIC values, indicating improved fit without undue complexity. The performance enhancement is most pronounced for the stretched exponential model, which achieves the highest $R^2$ (approximately 0.68) and the lowest RMSE (about 1.41), together with a markedly lower AIC than the conventional formulation. These metrics demonstrate that the modified decline-curve formulations provide a consistently superior representation of the observed temperature decline behavior while maintaining favorable complexity-adjusted model performance.

Although the modified decline-curve formulations introduce an additional fitting parameter, the observed increases in $R^2$ and reductions in RMSE are not considered in isolation when assessing model performance. For this reason, model comparison is based primarily on the Akaike Information Criterion (AIC), which explicitly accounts for model complexity by penalizing the inclusion of extra parameters. The consistently lower AIC values obtained for all modified formulations demonstrate that the observed improvements arise from a more favorable trade-off between goodness of fit and model simplicity, rather than from overfitting the temperature data. Furthermore, the inclusion of the equilibrium-temperature term is physically grounded, as it enforces finite long-term behavior that is consistent with geothermal heat-transfer processes. Importantly, when the equilibrium temperature is set to zero, the modified equations reduce exactly to the conventional Arps decline forms, ensuring backward compatibility with classical petroleum-based decline analysis.

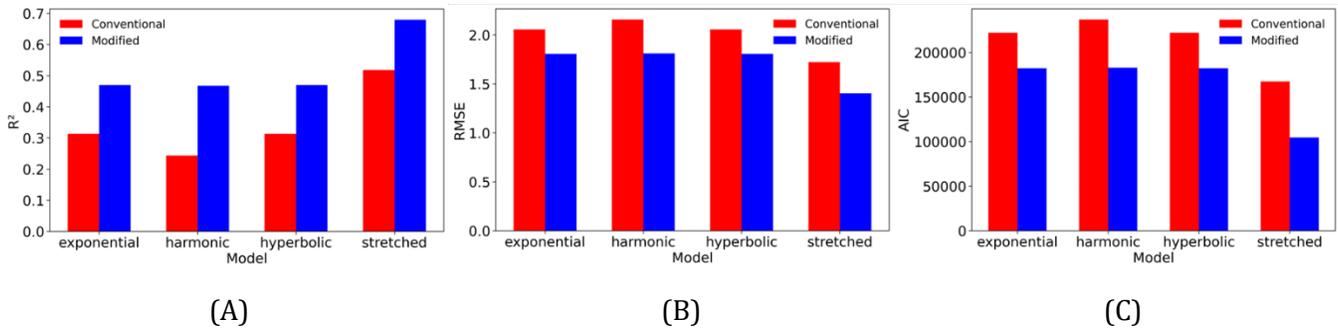

*Figure 12. Statistical performance of conventional and modified decline-curve models: (A) coefficient of determination, (B) RMSE, and (C) Akaike Information Criterion.*

## 6. Results and Analysis

This section presents the main findings from the full thermo-hydro-mechanical dataset. The results are arranged to show how each modeling approach captures production-temperature behavior and how well the methods generalize across the design space. The extended decline-curve models are first evaluated on the temperature histories. The equation-informed neural network is then examined for its ability to reproduce full trajectories from design inputs. Finally, the Gaussian Process and XGBoost regression models are analyzed for their accuracy and their uncertainty-aware forecasts across multiple time horizons.

### 6.1. Geothermal Decline Curve Analysis

The analysis evaluated 110 cases of time-temperature decline using four empirical decline forms and selected the best representation per case using AIC. Across all best-fit cases, model fits were exceptionally tight, with a median coefficient of determination of 0.9999, a median root-mean-square error of 0.071 °C, and a median mean absolute error of 0.048 °C (Table 3). Selection by the AIC criterion favored the stretched-exponential form in 66 of 110 cases (60.0%), the hyperbolic form in 41 cases (37.3%), and the harmonic form in 3 cases (2.7%); the simple exponential alternative was not selected as the most plausible description (Figure 13). These outcomes indicate that most temperature-decline trajectories exhibit either mild deviation from a pure exponential response, consistent with a stretched relaxation process, or curvature captured by the hyperbolic equation.

*Table 3.* *Statistics of the best-fit models across 110 cases: mean, median, and standard deviation for $R^2$, RMSE, and MAE.*

| Statistic | $R^2$ | RMSE, °C | MAE, °C |
|---|---|---|---|
| mean | 0.9998 | 0.116 | 0.084 |
| median | 0.9999 | 0.071 | 0.048 |
| Standard deviation | 0.0003 | 0.124 | 0.094 |

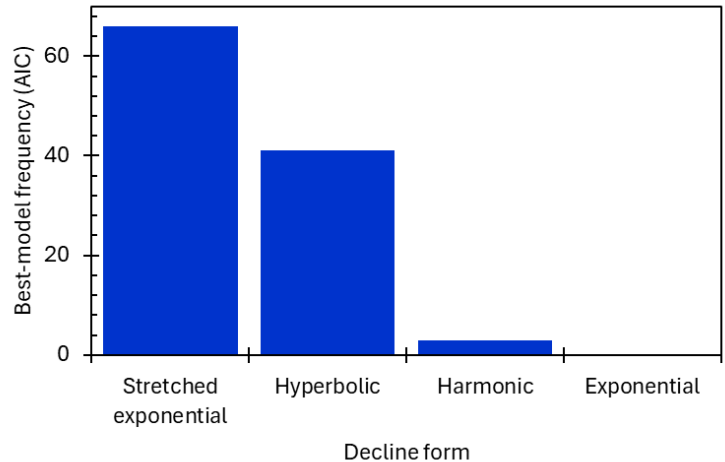

*Figure 13.* *AIC-best frequency across 110 cases; stretched-exponential dominates, followed by hyperbolic.*

Goodness-of-fit metrics clustered at low error levels for the selected models, supporting the reliability of the inferred characteristic parameters. Among the selected fits, the stretched-exponential cases achieved the lowest typical errors and strongest parsimony-adjusted support, with median RMSE ≈ 0.043, and median MAE ≈ 0.032 (Table 4). Hyperbolic selections also performed strongly, with median RMSE ≈ 0.089, and median MAE ≈ 0.069. The three harmonic selections showed median RMSE ≈ 0.068 and median MAE ≈ 0.0385, but the small sample indicates these were limited to specific geometries of decline. Taken together, the error distributions demonstrate that DCA models can reproduce the observed temperature trajectories within a few hundredths of a degree, with a consistent advantage for the stretched exponential form. RMSE for the 110 best-fit cases is illustrated in Figure 14.

*Table 4.* *Per-model performance statistics for the 110 best-fit cases: mean, median, and standard deviation of $R^2$, RMSE, and MAE.*

| Model | AIC-best cases | $R^2$ | | | RMSE, °C | | | MAE, °C | | |
|---|---|---|---|---|---|---|---|---|---|---|
| | | mean | median | std | mean | median | std | mean | median | std |
| Harmonic | 3 | 0.9999 | 0.9999 | 9.0E-06 | 0.0677 | 0.0676 | 0.0008 | 0.0386 | 0.0385 | 0.0002 |
| Hyperbolic | 41 | 0.9998 | 0.9999 | 0.0003 | 0.1250 | 0.0890 | 0.0982 | 0.0971 | 0.0693 | 0.0883 |
| Stretched Exp. | 66 | 0.9998 | 0.9999 | 0.0003 | 0.1125 | 0.0429 | 0.1397 | 0.0786 | 0.0318 | 0.0983 |

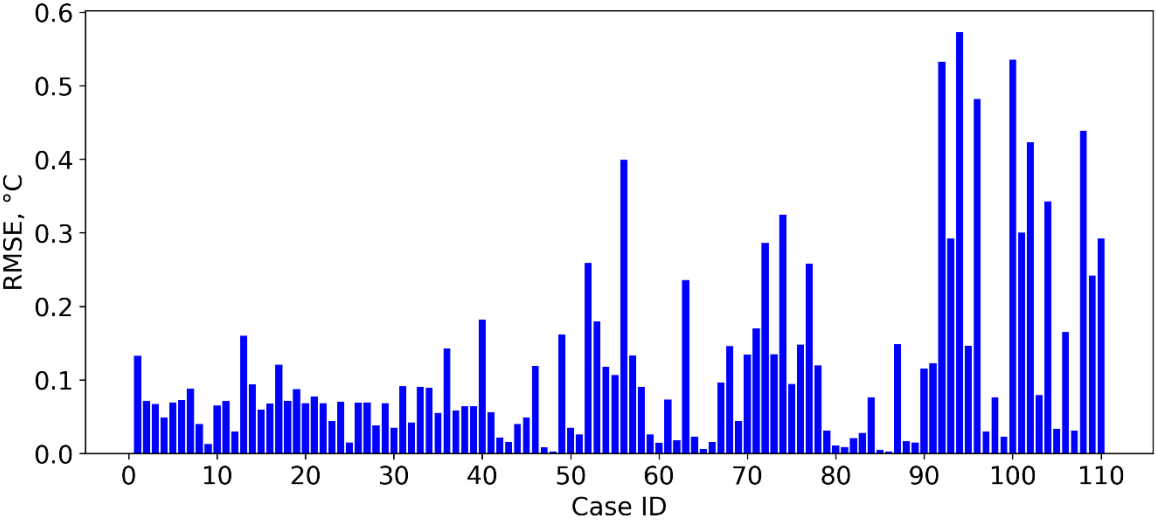

*Figure 14.* *RMSE for the 110 AIC-best-fit cases using DCA.*

Overall, temperature-decline curve across the dataset is best described by a stretched relaxation process in the majority of cases (Figure 15). Error magnitudes are uniformly small, parameter estimates are stable and interpretable, and selection preferences are consistent with trajectories that depart modestly from single-rate equilibration. The recommended presentation emphasizes the compactness of fit residuals, and representative overlays of observed and selected curves.

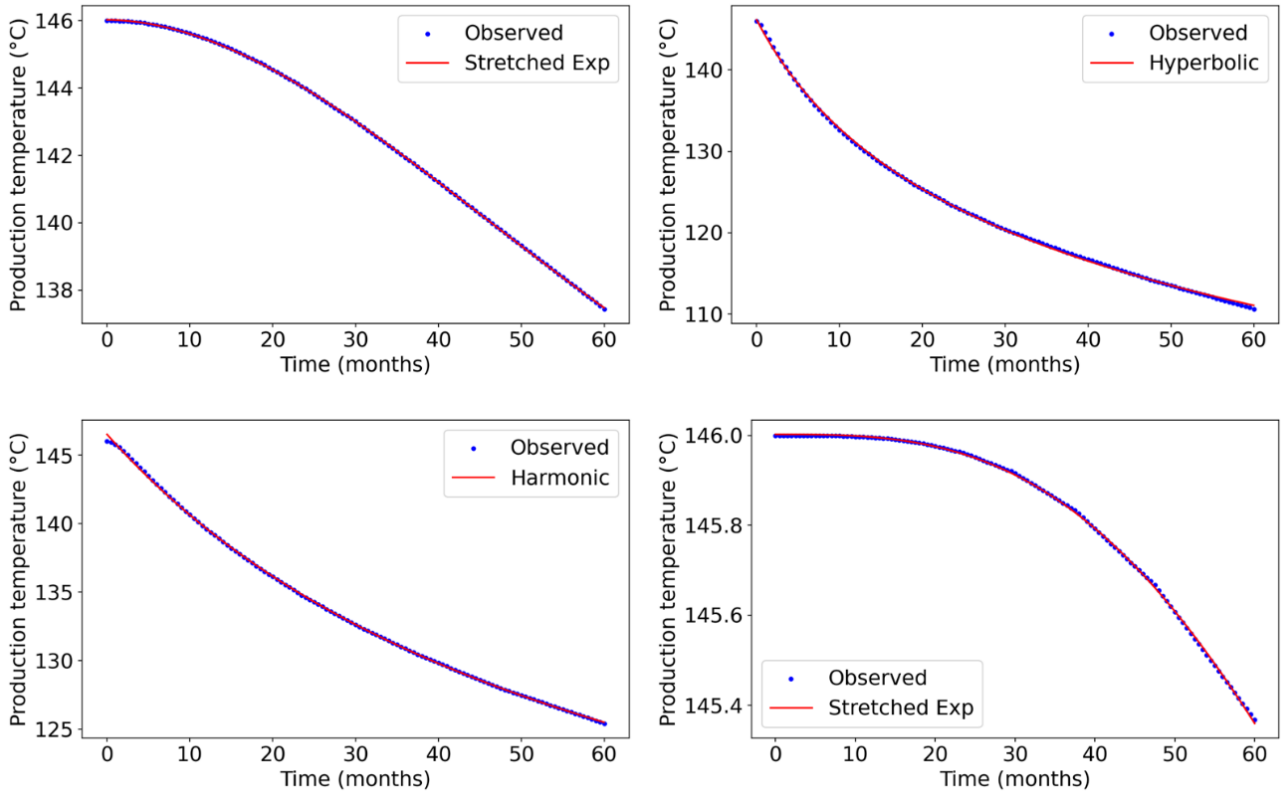

*Figure 15.* Representative DCA fits illustrating early-time curvature capture by stretched-exponential and flexible tapering by hyperbolic.

### 6.2. Equation-Informed Neural Network (EINN)

The model was trained using 440 fitted decline cases representing four functional forms (exponential, harmonic, hyperbolic, and stretched exponential) across 110 reservoir scenarios, as described in Section 6.1. The fitting quality of these cases was consistently high, with $R^2$ values ranging from 0.85 to 1.00 and an average of 0.99, indicating strong agreement between DCA fits and reference temperature histories. A stratified 20% of the cases (88 in total, 22 per decline form) were reserved for testing, while the remaining 80% were used for model development with 5-fold cross-validation.

Cross-validation on the training-validation pool corroborated these findings. Aggregated across folds, the overall validation errors were MAE 5.3 °C and RMSE 7.292 °C (Table 5). By form, the aggregated validation means were: harmonic MAE 5.15 °C, RMSE 7.066 °C; stretched-exponential MAE 5.259 °C, RMSE 7.233 °C; exponential MAE 5.045 °C, RMSE 6.837 °C; hyperbolic MAE 5.745 °C, RMSE 7.923 °C. The box plot distributions of MAE and RMSE over all cross-validation folds for the decline forms are illustrated in Figure 16.

Figure 17 presents parity plots for the EINN predictions versus the reference temperatures on the training set at four representative horizons (15, 30, 45, and 60 months). Across all horizons, points cluster tightly around the 1:1 line, indicating strong agreement and minimal systematic bias over the full temperature range. Predictive accuracy remains high at early and mid horizons (15–30 months), with $R^2 = 0.98$ and RMSE ≈ 2.82-2.88 °C, and shows only modest degradation at later times as variability increases ($R^2 = 0.97$, RMSE = 3.89 °C at 45 months; $R^2 = 0.96$, RMSE = 4.94 °C at 60 months). The gradual increase in scatter at 60 months suggests that long-horizon errors accumulate slightly, while overall calibration remains strong and the dominant trend is preserved.

*Table 5. Cross-validated out-of-fold performance of the EINN model by decline form. For each form, the table reports the mean ± standard deviation of MAE, and RMSE across all folds.*

| Form / Model | N | MAE (°C) (mean ± SD) | RMSE (°C) (mean ± SD) | Folds |
| --- | --- | --- | --- | --- |
| Exponential | 88 | 5.045 ± 1.176 | 6.837 ± 1.465 | 5 |
| Harmonic | 88 | 5.150 ± 1.234 | 7.066 ± 1.505 | 5 |
| Hyperbolic | 88 | 5.745 ± 1.299 | 7.923 ± 1.609 | 5 |
| Stretched Exp | 88 | 5.259 ± 0.912 | 7.233 ± 1.076 | 5 |
| All | 352 | 5.300 ± 1.074 | 7.292 ± 1.322 | 5 |

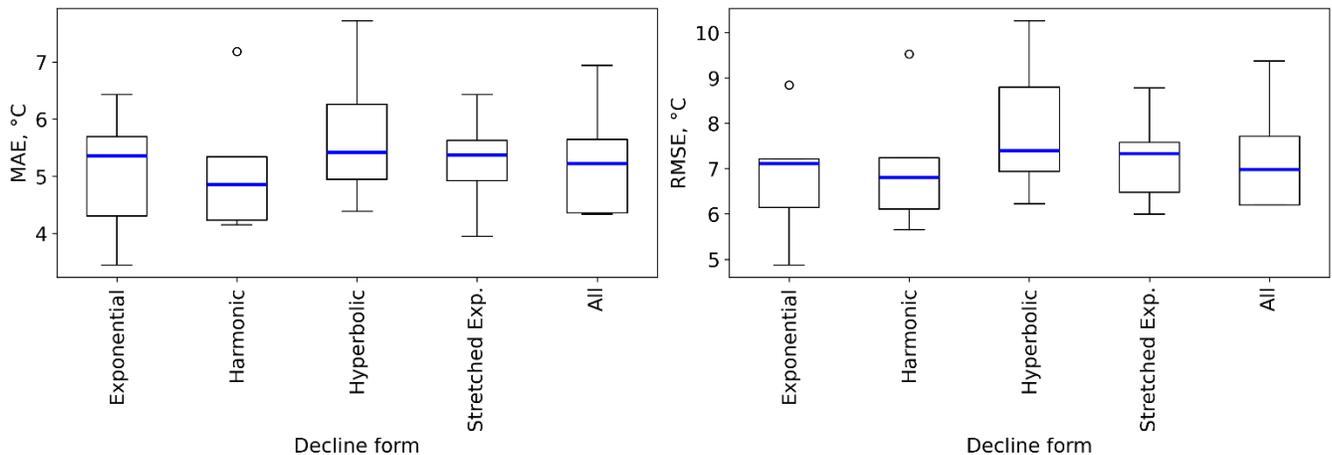

*Figure 16. Predictive performance of the EINN model across decline forms. Box plots show the distribution of MAE and RMSE over all cross-validation folds for each form. The All category represents pooled results across all forms.*

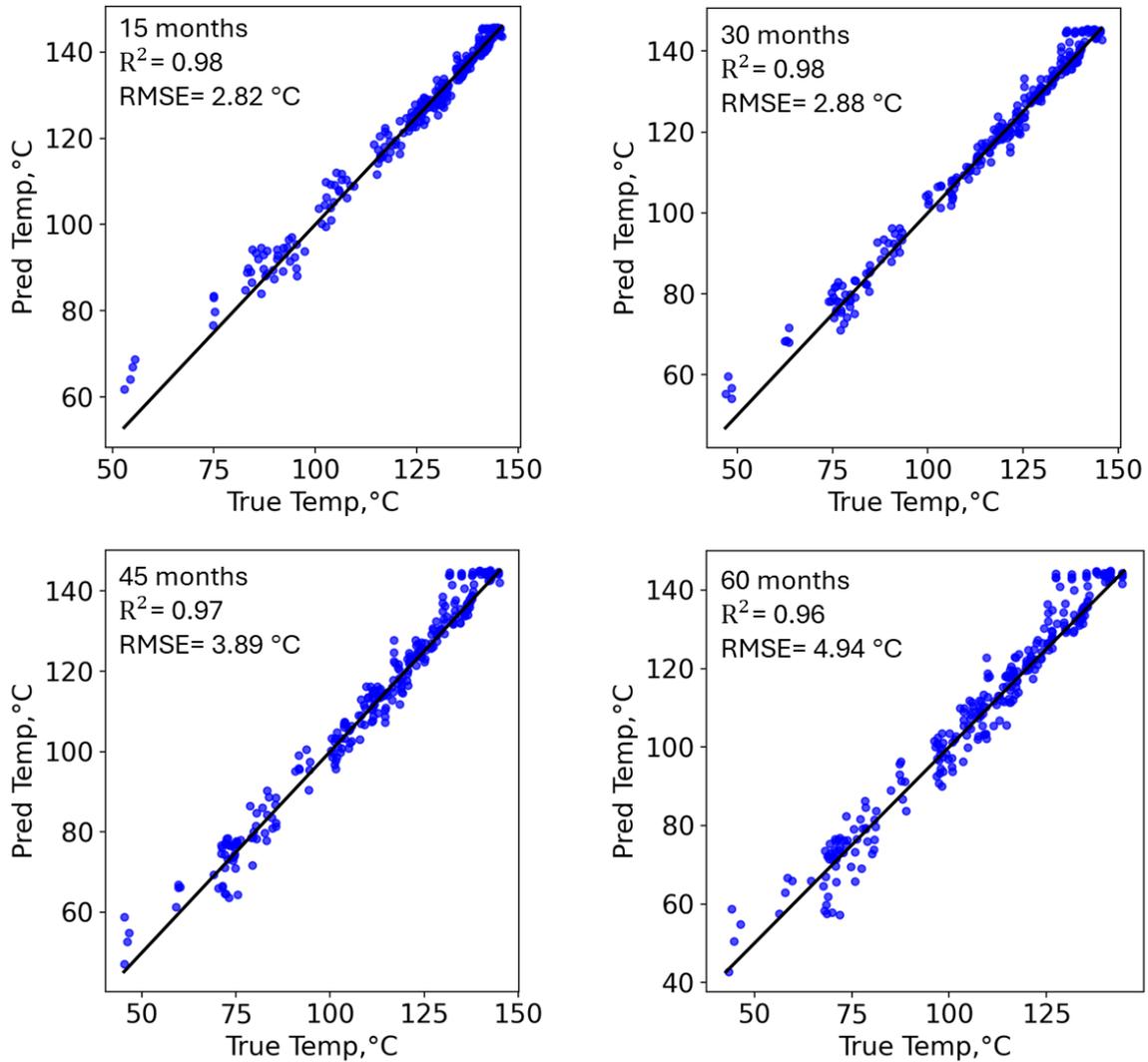

*Figure 17. Parity plots for EINN-predicted vs. true temperatures on the training set at four representative time horizons (15, 30, 45, and 60 months). Each panel shows the 1:1 reference line with horizon-specific $R^2$ and RMSE.*

Across the independent test set, the method achieved strong agreement with the reference curves (Table 6). The overall mean absolute error was 3.06 °C with a corresponding root-mean-square error of 4.49 °C. Performance was consistent across decline forms: stretched-exponential exhibited the lowest typical errors (MAE 2.586 °C, RMSE 3.423 °C), followed by harmonic (MAE 2.914 °C, RMSE 4.556 °C). Exponential and hyperbolic showed higher dispersion but still favorable agreement (exponential: MAE 3.248 °C, RMSE 4.684 °C; hyperbolic: MAE 3.489 °C, RMSE 5.117 °C). These results indicate reliable reproduction of reference temperature trajectories over the 0–60-month horizon, with particularly tight alignment for the stretched exponential and harmonic forms.

*Table 6. Hold-out test performance of the EINN model by decline form. For each form, the table reports the mean ± standard deviation of $R^2$, MAE, and RMSE across all test cases.*

| Form / Model | N | MAE (°C) (mean ± SD) | RMSE (°C) (mean ± SD) |
|---|---|---|---|
| Exponential | 22 | 3.248 ± 2.689 | 4.684 ± 4.377 |
| Harmonic | 22 | 2.914 ± 2.932 | 4.556 ± 4.935 |
| Hyperbolic | 22 | 3.489 ± 2.678 | 5.117 ± 3.902 |
| Stretched Exp | 22 | 2.586 ± 1.726 | 3.423 ± 2.498 |
| All | 88 | 3.059 ± 2.571 | 4.489 ± 4.205 |

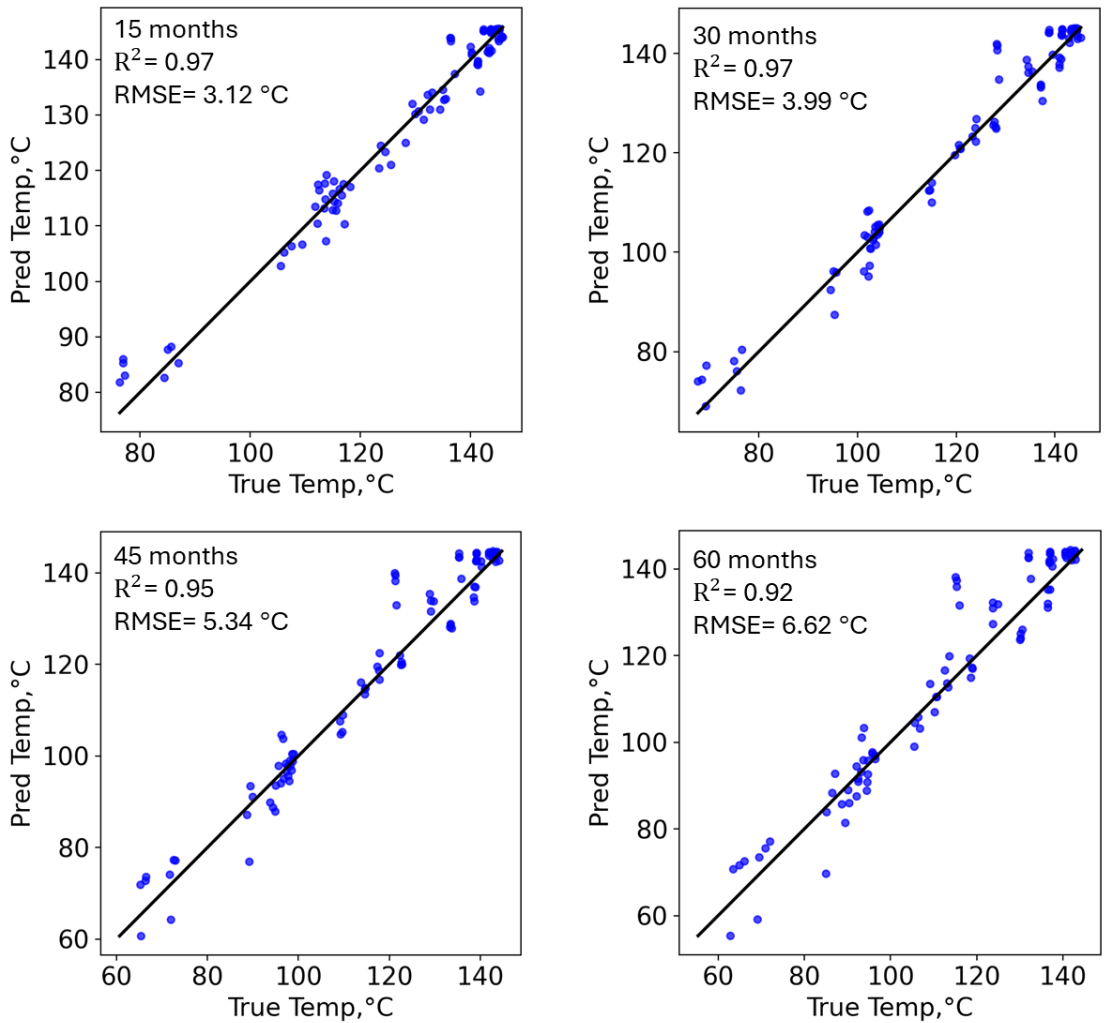

*Figure 18. Parity plots for EINN-predicted vs. true temperatures on the 22 hold-out cases at four representative time horizons (15, 30, 45, and 60 months). Each panel shows the 1:1 reference line with horizon-specific $R^2$ and RMSE, demonstrating strong predictive agreement across short- and long-term forecasts.*

Figure 18 presents parity plots comparing EINN-predicted temperatures with the true values for the 22 hold-out cases at four forecast horizons (15, 30, 45, and 60 months). In each panel, the points cluster tightly around the 1:1 reference line, indicating strong calibration between predicted and observed temperatures. The goodness-of-fit remains high across all horizons, with $R^2$ values ranging from 0.92 to 0.97 and RMSE values between 3.12 °C and 6.62 °C. Although scatter increases at the longest horizon, the overall alignment demonstrates that the model maintains reliable predictive accuracy from early to late stages of thermal decline, confirming its ability to generalize beyond the training data.

Figure 19 compares the DCA temperature-decline curves with the corresponding EINN-predicted trajectories for several representative hold-out cases over the full 3–60 month window. In all examples, the ML-generated curves (red) closely overlap the reference DCA curves (blue), matching both the early steep cooling phase and the gradual late-time taper. Only minor deviations are observed, indicating that the equation-informed network is able to reproduce the full temporal shape of the decline rather than fitting isolated points. The visual agreement across a range of decline curves confirms that the model preserves physical curve structure while providing accurate, data-driven predictions suitable for surrogate forecasting.

Overall, the results demonstrate that the proposed approach reproduces reference temperature decline curve with high fidelity, delivering around 3 °C typical absolute errors for the best-performing forms and maintaining strong agreement across the entire test set, supporting use in prospective applications where rapid, reliable curve characterization is required.

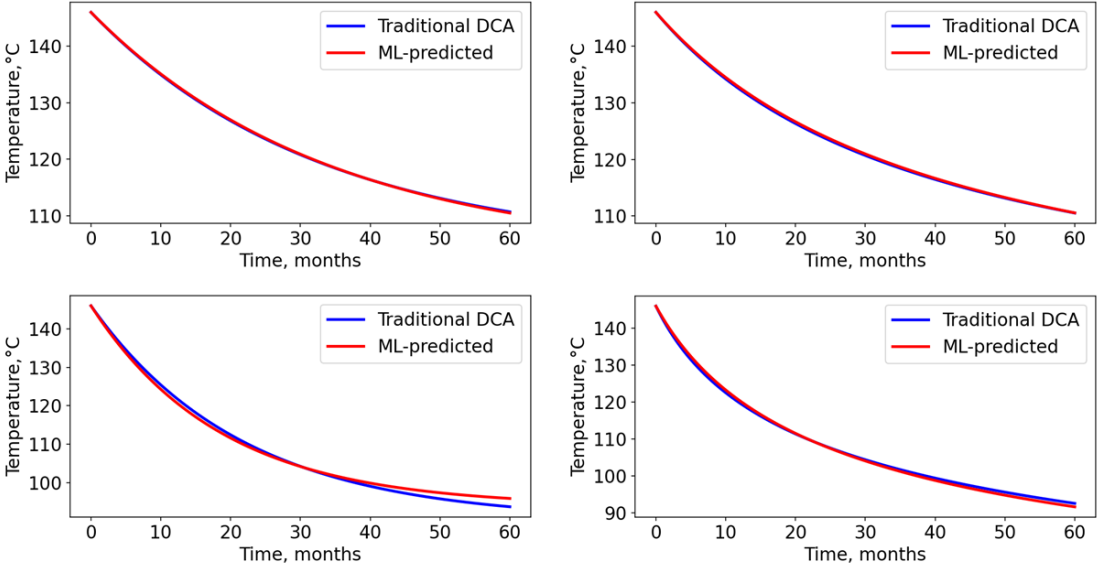

*Figure 19. Observed (DCA-based) vs EINN-predicted temperature trajectories for representative hold-out test cases over 3–60 months.*

## 6.3. Gaussian Process Regression

The analysis used a group-aware design to estimate generalization across distinct cases. A single, unseen hold-out set comprising 22 cases was reserved strictly by case identity, while remaining cases were used for out-of-fold assessment via grouped cross-validation. Targets were monthly temperatures at 3-month increments from 3 to 60 months (20 horizons). This design mitigates leakage between related observations and permits two complementary views of performance: cross-validated training out-of-fold (OOF) metrics and a once-only evaluation on the hold-out set.

On the cross-validated training assessment, the macro performance across all 20 horizons yielded $R^2 = 0.934$ (SD 0.030), RMSE = 3.67 (SD 0.26) °C, and MAE = 2.57 (SD 0.22) °C, summarizing fold-level means aggregated over targets (Table 7). Performance improved monotonically with forecast horizon, from $R^2 \approx 0.819$ at 3 months to $R^2 \approx 0.951$ at 60 months, with a corresponding gradual reduction in error magnitudes across horizons. These OOF results indicate stable fit quality within training data partitions while retaining group separation.

*Table 7. Cross-validated out-of-fold (OOF) performance of the GPR surrogate across sampled prediction horizons (3–60 months). Values are reported as mean ± standard deviation over grouped folds for $R^2$, RMSE, and MAE. The Macro row summarizes the average performance across all horizons.*

| Horizon (months) | $R^2$ (mean ± SD) | RMSE (°C) (mean ± SD) | MAE (°C) (mean ± SD) |
|---|---|---|---|
| 3 | 0.819 ± 0.290 | 2.679 ± 1.186 | 1.771 ± 0.615 |
| 9 | 0.931 ± 0.052 | 3.487 ± 1.232 | 2.363 ± 0.678 |
| 15 | 0.936 ± 0.063 | 3.765 ± 1.090 | 2.591 ± 0.516 |
| 21 | 0.940 ± 0.075 | 3.794 ± 1.171 | 2.676 ± 0.605 |
| 27 | 0.944 ± 0.075 | 3.739 ± 1.245 | 2.629 ± 0.674 |
| 33 | 0.946 ± 0.074 | 3.719 ± 1.298 | 2.597 ± 0.741 |
| 39 | 0.948 ± 0.074 | 3.691 ± 1.346 | 2.595 ± 0.811 |
| 45 | 0.946 ± 0.074 | 3.871 ± 1.516 | 2.763 ± 1.087 |
| 51 | 0.948 ± 0.072 | 3.815 ± 1.487 | 2.743 ± 1.048 |
| 57 | 0.950 ± 0.070 | 3.741 ± 1.447 | 2.697 ± 0.971 |
| 60 | 0.951 ± 0.069 | 3.712 ± 1.428 | 2.680 ± 0.932 |
| Macro | 0.934 ± 0.030 | 3.669 ± 0.262 | 2.573 ± 0.220 |

Figure 20 shows parity plots comparing GPR-predicted and true temperatures for the training cases (based on out-of-fold) at four representative horizons (15, 30, 45, and 60 months). The predictions align closely with the 1:1 reference line at all horizons. Model performance is consistent over time, with coefficients of determination remaining high ($R^2$ = 0.97-0.98) and RMSE values confined to a narrow band of approximately 3.9-4.2 °C. No pronounced horizon-dependent degradation is observed; instead, scatter remains comparable from early to late times, suggesting that the GPR surrogate captures both short-term and longer-term thermal behavior with stable accuracy. Minor dispersion at lower temperatures reflects case-specific variability rather than systematic error, while overall trends are well preserved across all horizons.

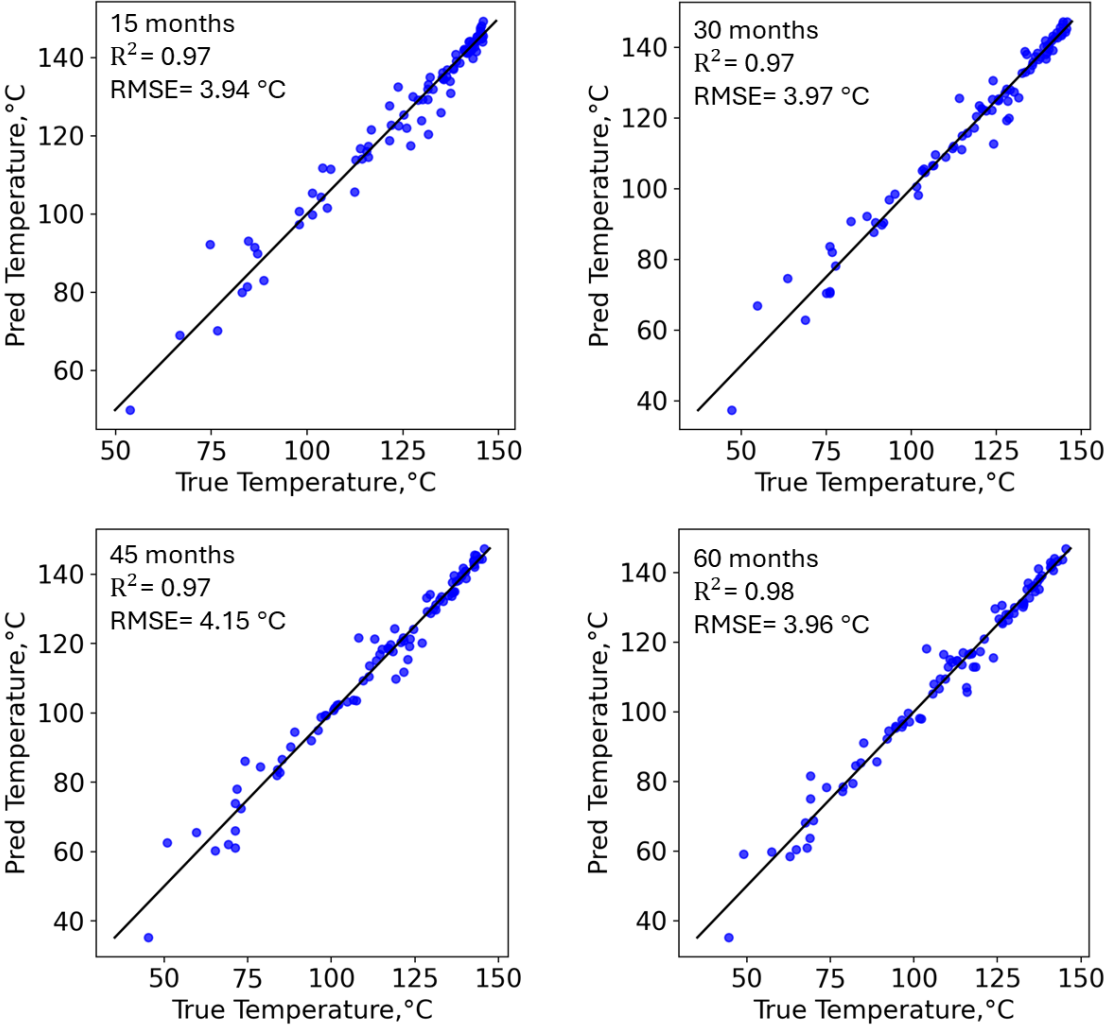

*Figure 20. Parity plots for predicted vs. true temperatures on the 88 training cases at four representative time horizons (15, 30, 45, and 60 months) using the GPR surrogate. Each panel shows the 1:1 reference line with horizon-specific $R^2$ and RMSE.*

On the hold-out evaluation, macro performance across the 20 horizons totaled 440 target evaluations and reached $R^2$ = 0.965 (SD across horizons 0.013), RMSE = 3.39 (SD 0.24) °C, and MAE = 2.34 (SD 0.20) °C (Table 8). Horizon-specific performance was uniformly strong: the lowest $R^2$ occurred at 3 months ($R^2$ = 0.933), rising steadily to the highest value at 60 months ($R^2$ = 0.976). Errors remained in a narrow band, with RMSE ranging from 2.45 °C (3 months) to a modest peak of 3.67 °C around 21 months, then stabilizing near 3.38 °C toward the later horizons. MAE showed a similar profile, averaging 2.34 °C across horizons. Taken together, the hold-out metrics are slightly stronger than the OOF estimates, consistent with stable generalization.

Inspection of per-target dispersion in the hold-out residuals (SD of absolute errors) shows broad consistency across horizons, supporting homogeneity of predictive reliability over time. Because targets are measured on the same scale, the near-parallel behavior of RMSE and MAE across horizons suggests that error distributions are not dominated by outliers. This pattern, together with steadily increasing $R^2$ values at later months, indicates that predictive signal grows with forecast horizon while absolute errors remain well-bounded. The error profiles (RMSE and MAE) across time horizons for the GPR surrogate are shown in Figure 21 for both out-of-fold validation and hold-out test sets across 20 monthly horizons (3–60 months).

*Table 8. Hold-out test performance of the GPR surrogate across sampled prediction horizons (3–60 months). Metrics are reported for all 22 test cases at each horizon as mean ± standard deviation for $R^2$, RMSE, and MAE. The Macro row summarizes the average performance across all horizons.*

| Horizon (months) | N | $R^2$ (mean ± SD) | RMSE (°C) (mean ± SD) | MAE (°C) (mean ± SD) |
|---|---|---|---|---|
| 3 | 22 | 0.933 | 2.446 ± 2.327 | 1.803 ± 1.653 |
| 9 | 22 | 0.947 | 3.381 ± 3.376 | 2.397 ± 2.384 |
| 15 | 22 | 0.955 | 3.613 ± 3.605 | 2.729 ± 2.368 |
| 21 | 22 | 0.961 | 3.667 ± 3.650 | 2.584 ± 2.602 |
| 27 | 22 | 0.971 | 3.331 ± 3.325 | 2.345 ± 2.366 |
| 33 | 22 | 0.972 | 3.388 ± 3.367 | 2.344 ± 2.446 |
| 39 | 22 | 0.973 | 3.419 ± 3.376 | 2.311 ± 2.519 |
| 45 | 22 | 0.975 | 3.413 ± 3.349 | 2.266 ± 2.553 |
| 51 | 22 | 0.975 | 3.403 ± 3.325 | 2.234 ± 2.567 |
| 57 | 22 | 0.976 | 3.384 ± 3.298 | 2.193 ± 2.578 |
| 60 | 22 | 0.976 | 3.381 ± 3.290 | 2.172 ± 2.592 |
| Macro | 440 | 0.965 ± 0.013 | 3.386 ± 0.242 | 2.337 ± 0.202 |

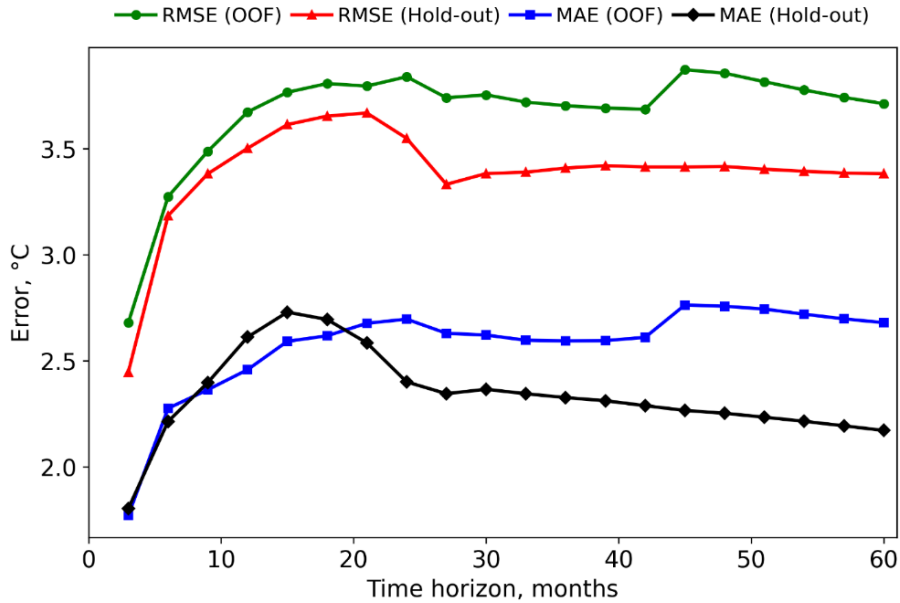

*Figure 21. Error profiles across time horizons for the GPR surrogate. RMSE and MAE are shown for both out-of-fold (OOF) validation and hold-out test sets across 20 monthly horizons (3–60 months). Error magnitudes remain stable over time, confirming consistent predictive performance.*

Figure 22 compares the GPR-predicted temperatures with the true hold-out values at four representative forecast horizons (15, 30, 45, and 60 months). Across all horizons, the points cluster tightly around the 1:1 line, indicating strong calibration and low systematic bias. The coefficient of determination increases slightly with forecast length ($R^2$ = 0.955:0.976), while RMSE remains stable at approximately 3.4 °C, confirming that predictive accuracy does not deteriorate at later horizons. This consistency shows that the surrogate model retains reliable generalization even when extrapolating further into the production period. Figure 23 illustrates the full temperature-decline trajectories for representative hold-out wells, comparing the true observations with the GPR predictions over all 20 forecast horizons (3-60 months). The predicted curves closely follow the observed behavior throughout the production period, and the narrow 95 % predictive intervals indicate low epistemic uncertainty.

These results support accurate and well-calibrated predictions across all target months, with consistent absolute errors and increasing explained variance at longer horizons. The alignment between OOF and hold-out performance, together with the scatter and overlay diagnostics, indicates robust generalization to unseen cases while preserving interpretability at the individual-trajectory level.

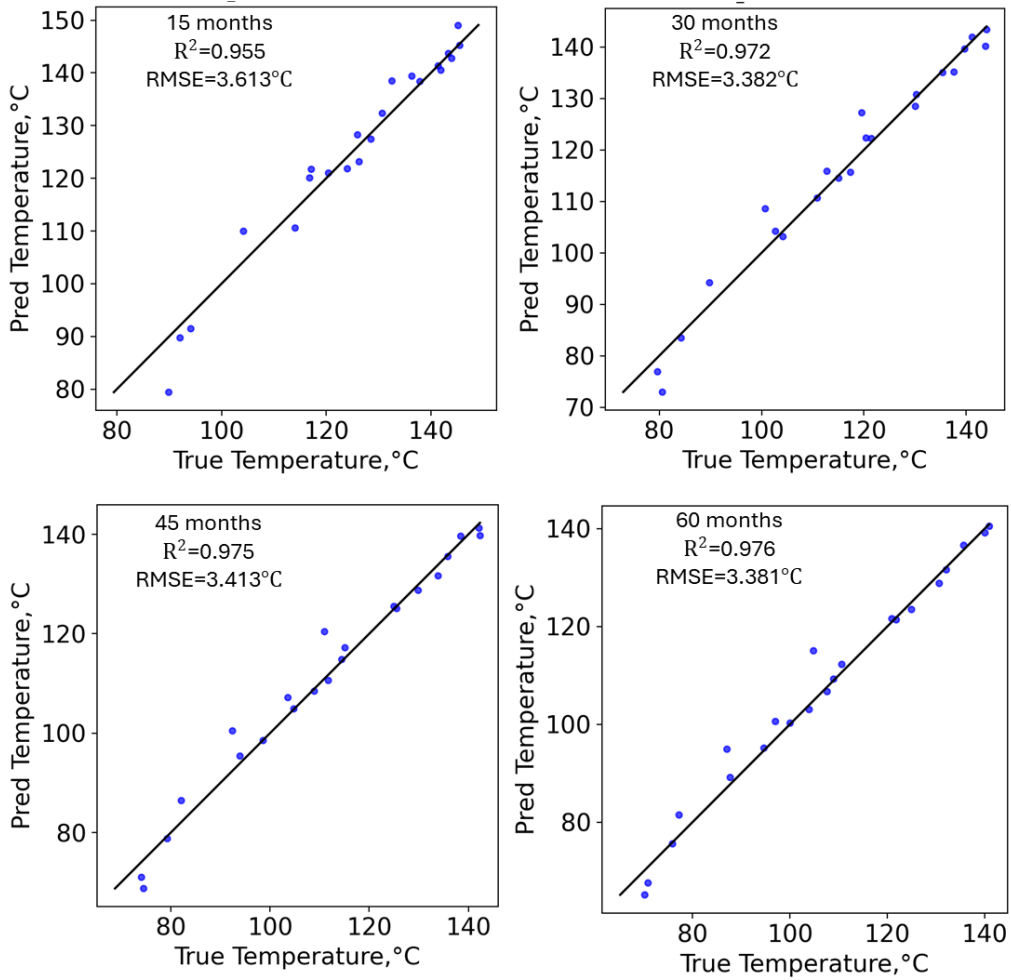

*Figure 22.* Predicted vs observed temperature for the 22 hold-out cases at four representative horizons (15, 30, 45, and 60 months) using the GPR surrogate. Each panel shows the one-to-one reference line, with horizon-specific $R^2$ and RMSE. The model maintains strong calibration across early and late prediction horizons.

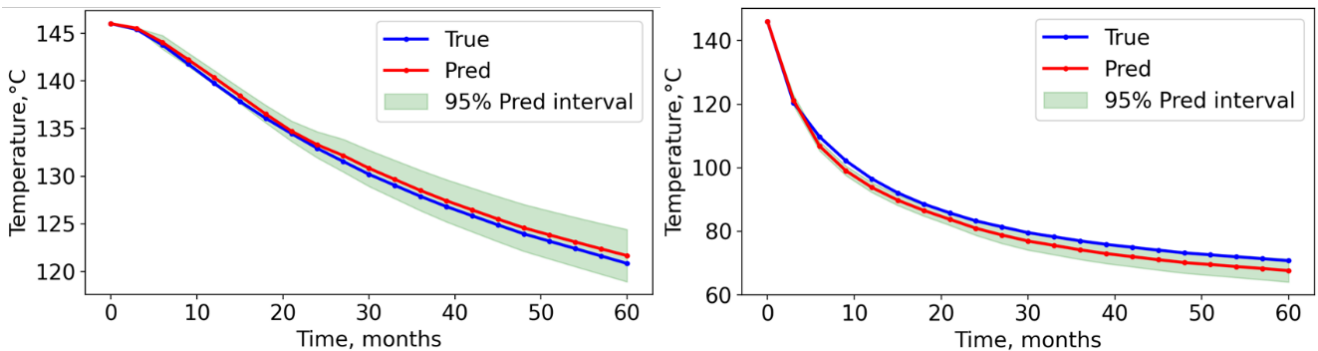

*Figure 23.* Observed and GPR-predicted temperature evolution for selected hold-out cases. Solid lines show true and predicted temperatures across 20 horizons (3–60 months), with shaded bands indicating 95% predictive/confidence intervals.

### 6.4. XGBoost Surrogate

This section presents the performance of the XGBoost model. The dataset was partitioned into two disjoint subsets at the THM case level, ensuring that all observations belonging to the same case were assigned exclusively to either training or testing. A 20% proportion of cases (22 cases) was reserved as a completely unseen test set, while the remaining 88 cases formed the training set. This split guarantees that reported results reflect model behavior on unseen cases.

Across all times, both error metrics (RMSE and MAE) increase at early horizons and then stabilize (Figure 24). From the earliest horizon (3 months) to roughly 12–18 months, RMSE and MAE rise for both training and testing, indicating that predictive accuracy degrades as the forecast horizon extends from short to annual scales. After 20 months approximately, the curves flatten, showing that additional increases in time horizon produce comparatively small changes in error.

Training errors (computed across multiple cross-validation folds and averaged) are consistently higher than testing errors at every horizon. Training RMSE starts near 7.2 °C and increases to a maximum of 9.5 °C near 24 months, then slowly trends downward to about 9.0 °C by 60 months. Training MAE begins near 4.7 °C, climbs to 7.2 °C by 24 months, and remains close to 7.1 °C through the longest horizons. This pattern indicates that, once the horizon reaches about two years, the typical magnitude of training errors becomes largely time-invariant, with only minor fluctuations.

Testing performance shows the same rise-then-plateau behavior but at lower error levels. Test RMSE increases from roughly 5.5 °C at the shortest horizon to about 7.2 °C around 18 months, followed by a stabilization and slight improvement toward 6.9 °C by 60 months. Test MAE rises from approximately 3.7 °C to 5.8 °C around 21-33 months, then gradually decreases to about 5.4 °C at the longest horizons. The small late-horizon reduction in test MAE (and mild softening of test RMSE) suggests that errors saturate and can even improve at multi-year horizons.

As the gap between training and testing curves does not widen with horizon, it reflects a stable offset rather than horizon-dependent divergence. Overall, the results indicate three key outcome features: (1) most degradation with horizon occurs early (up to 20 months), (2) multi-year horizons exhibit a clear error plateau, and (3) aggregated training errors remain systematically higher than testing errors throughout, while testing errors remain relatively steady and slightly improved at the longest horizons.

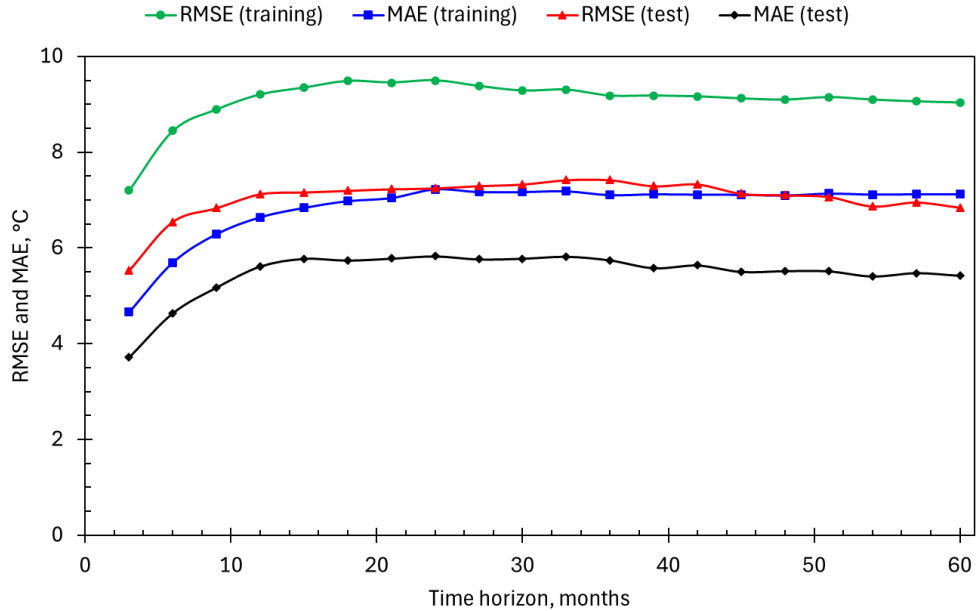

**Figure 24.** *Root mean square error (RMSE) and mean absolute error (MAE) as a function of prediction horizon (months) for the XGBoost models, shown for aggregated training results and for the test set.*

The training-data parity plots are given in Figure 25. The figure shows a consistently strong linear agreement between predicted and true temperature across all horizons (15, 30, 45, and 60 months). In every panel, most points cluster close to the 1:1 line, indicating that the model captures the dominant temperature signal and produces strongly correlated estimates. The reported coefficients of determination increase with the time horizon, from $R^2$ = 0.80 (15 months) to 0.84 (30 months), 0.86 (45 months), and 0.87 (60 months). This gradual rise indicates that, within the training set, explained variance improves modestly at longer horizons, with the 45–60 month horizons showing the tightest overall alignment. The error magnitudes are also stable and slightly improving with horizon. The reported RMSE decreases from 9.71 °C (15 months) to 9.55 °C (30 months), 9.37 °C (45 months), and 9.31 °C (60 months). The change is incremental rather than dramatic, implying that the typical training deviation remains on the order of 9-10 °C regardless of horizon, but with a small systematic reduction as the horizon extends.

The point patterns suggest that predictive fidelity is strongest in the higher-temperature regime (roughly 120–150 °C), where the cloud is dense and closely tracks the 1:1 line. In contrast, the lower-temperature region (roughly 50–90 °C) shows wider scatter and several visibly larger departures from the parity line. The accuracy appears more variable at lower true temperatures, while high temperatures are reproduced more consistently.

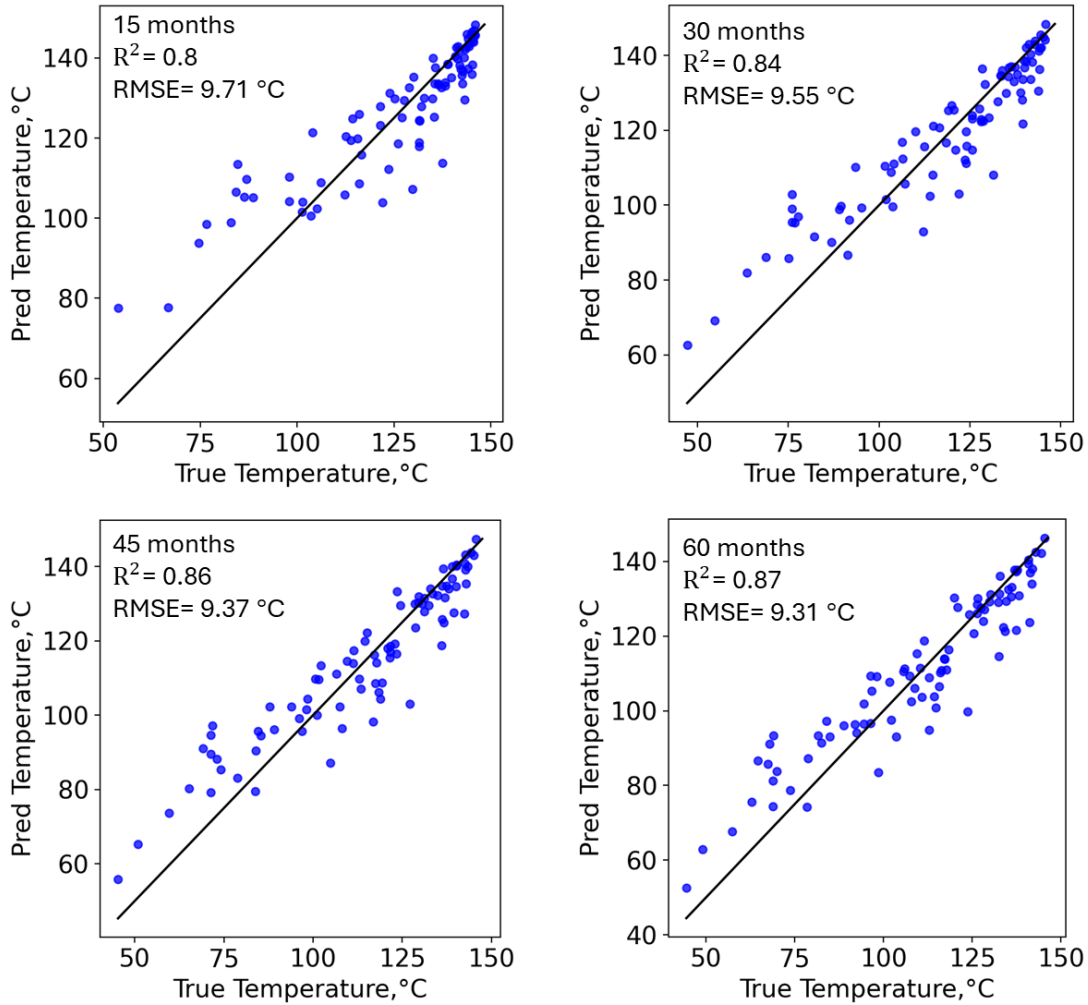

*Figure 25.* *Parity plots for the XGBoost training predictions at 15, 30, 45, and 60 months. Predicted temperatures are shown versus true temperatures. Each panel shows the one-to-one reference line, with horizon-specific $R^2$ and RMSE.*

Figure 26 shows hold-out testing scatter plots (Predicted vs. True temperature) for four horizons (15, 30, 45, 60 months). At 15 months, the relationship is strong ($R^2$ = 0.83, RMSE = 7.17 °C). Most points cluster near the 1:1 line at higher temperatures, indicating good accuracy in the upper range. The largest discrepancies appear among lower-to-mid true temperatures. At 30 months, performance improves in explained variance ($R^2$ = 0.87) while error magnitude remains similar (RMSE = 7.33 °C). The combination of higher $R^2$ with similar RMSE implies better overall ranking/linearity across samples. At 45 and 60 months, the agreement becomes more consistent. At 45 months ($R^2$ = 0.89, RMSE = 7.14 °C) and 60 months ($R^2$ = 0.90, RMSE = 6.85 °C), the scatter aligns more closely to the 1:1 line across most of the temperature range, and the largest residuals appear less frequent. The 60-month panel shows the best overall balance: the highest $R^2$ and the lowest RMSE among the four horizons, indicating that longer-horizon temperature predictions are captured with slightly better precision and more stable correspondence to the true values, while still retaining a small set of points that deviate materially from perfect agreement. Overall, these observations align with Figure 24 where early-horizon predictions are generally reliable but still show a few influential errors. Figure 27 shows representative test-set trajectories, comparing predicted and true temperature evolution for individual unseen cases across the full 0-60 month horizon. In each example, the model closely reproduces the overall cooling trend and temporal shape, with predictions generally tracking the true curve throughout time.

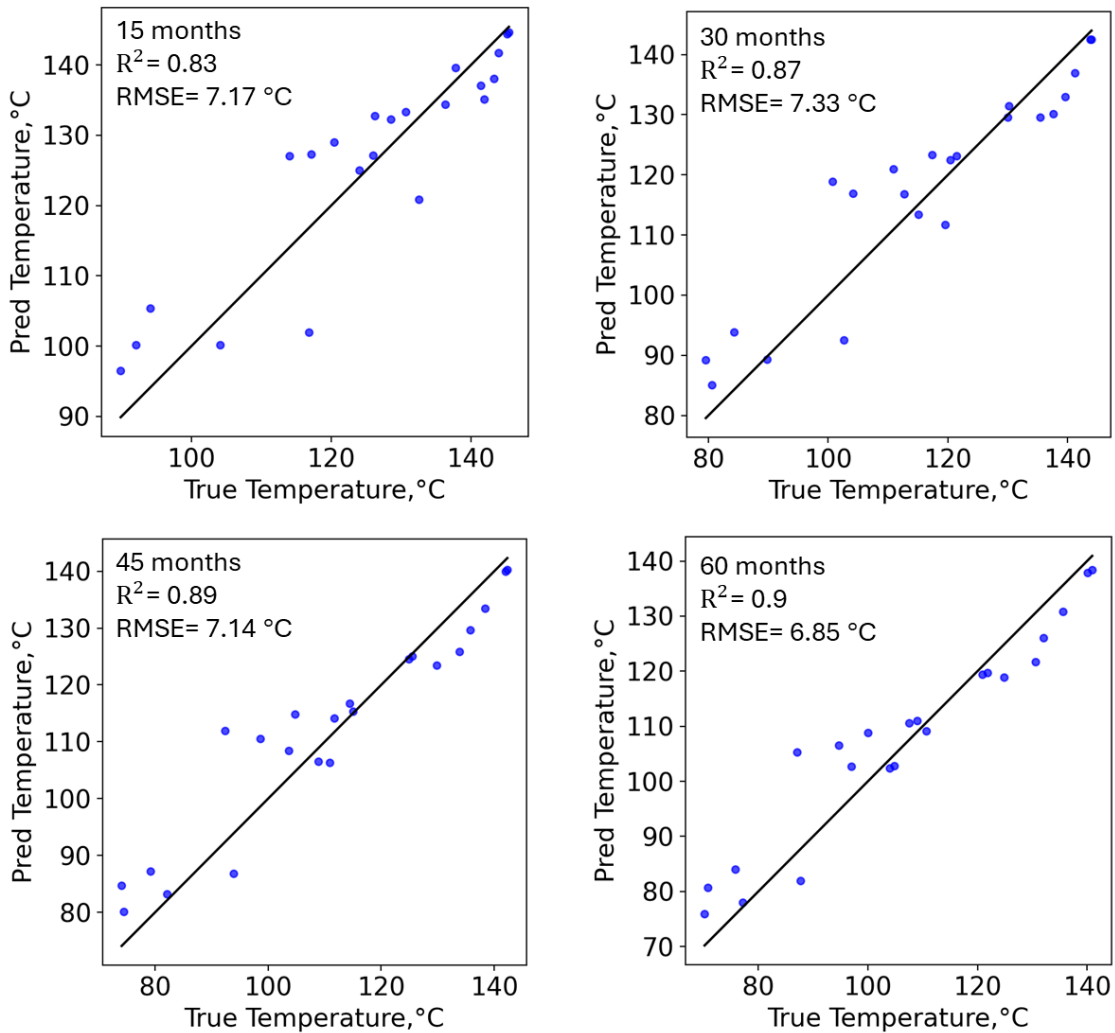

*Figure 26.* Parity plots for the XGBoost test predictions at 15, 30, 45, and 60 months. Predicted versus true temperatures are shown relative to the 1:1 line. Each panel shows the one-to-one reference line, with horizon-specific $R^2$ and RMSE.

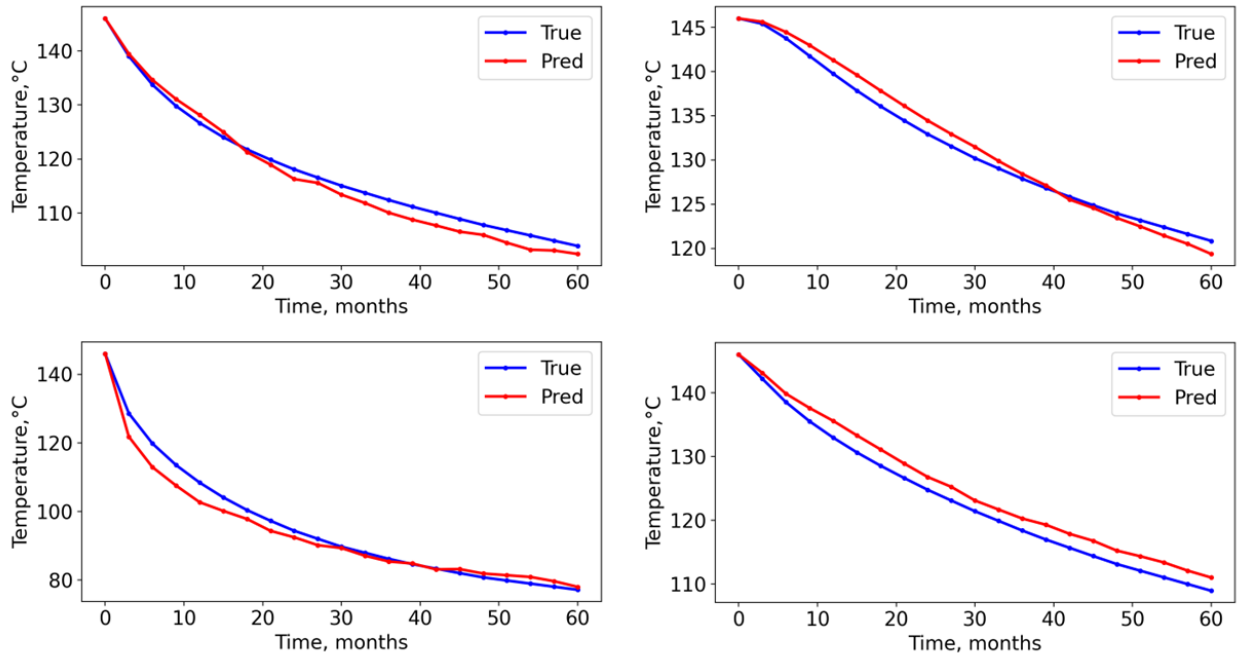

*Figure 27. Representative XGBoost test-set plots comparing true and predicted temperature trajectories for individual unseen cases over the 0–60 month horizon.*

## 7. Discussion

Because each thermo-hydro-mechanical simulation in the dataset requires solving fully coupled flow, heat transport, and mechanical deformation, a single high-fidelity run takes around 6-8 hours on a standard personal workstation (Intel i7-10750H, 16 GB RAM, Windows 11). This cost stems from the thermo-poroelastic coupling and the fine spatial and temporal resolution needed to capture fracture-controlled heat transport. In contrast, once trained, the equation-informed neural network and the Gaussian Process surrogate return full 0–60-month temperature forecasts almost instantly relative to the underlying simulator, allowing users to evaluate new configurations, conduct sensitivity studies, and explore design alternatives without incurring the heavy computational load of repeated THM simulations.

The equilibrium-temperature extension of the decline-curve formulations provides a physically consistent representation of geothermal cooling by enforcing finite long-term behavior governed by heat exchange with the surrounding formation. Validation against downhole temperature measurements from the Utah FORGE project demonstrates that the modified decline models capture observed geothermal temperature trends more reliably than conventional petroleum-based formulations, particularly at later times where asymptotic behavior becomes important. While the available field record is limited in duration, it represents the longest continuous publicly accessible downhole temperature dataset for an operating EGS production well suitable for decline-curve validation. In this context, the agreement between the modified decline curves and field observations provides important empirical support for the proposed formulation and establishes confidence in its use as a physics-consistent diagnostic and forecasting tool. Furthermore, when the equilibrium temperature is set to zero, the modified formulations reduce exactly to the conventional Arps decline equations, thereby preserving full backward compatibility with classical petroleum-based decline analysis. In addition, when applied to individual simulated cases, the modified models closely match the temperature data with very small errors.

The machine-learning models examined in this study serve complementary roles and differ primarily in how physical structure and uncertainty are incorporated. The equation-informed neural network (EINN) embeds geothermal-extended decline-curve equations directly within the learning architecture, enforcing physically consistent temperature-trajectory shapes and yielding reliable full-curve forecasts with typical test errors of MAE = 3.06 °C and RMSE = 4.49 °C. Gaussian Process Regression (GPR), implemented as a direct multi-horizon probabilistic surrogate, achieves the strongest overall predictive accuracy and stability across horizons (macro RMSE = 3.39 °C, MAE = 2.34 °C on the hold-out set) while providing explicit uncertainty quantification, making it well suited for rapid screening and risk-aware decision-making. In contrast, the direct XGBoost models, which do not embed physical structure or uncertainty, reproduce general trends but exhibit higher errors (around 7 °C RMSE), underscoring the value of incorporating physics or probabilistic structure. Despite the limited number of realizations used in model development, performance across all models remains stable on unseen cases, indicating that the dominant physical controls governing geothermal temperature decline are strongly expressed in the dataset and effectively captured by the surrogate formulations.

Recent work by Yan et al. [33] introduced the HyperReLU thermal-decline model and trained a fully connected neural network to regress its three decline parameters from thermos-hydro-mechanical simulation inputs, with physics incorporated through the analytical decline form and an equation-based regularization term in the loss function . The present study differs in two fundamental ways. First, the classical Arps decline family is generalized for geothermal temperature forecasting by introducing an equilibrium-temperature term consistent with Newton-type cooling, enforcing finite late-time thermal behavior while preserving exact reduction to the conventional Arps equations when the equilibrium term is set to zero . Second, rather than applying physics only as an external loss penalty on regressed parameters, the modified decline equations used in this study are embedded as differentiable computational layers inside an equation-informed neural network, ensuring that learning signals propagate through the analytical decline physics during training . In addition, the extended decline formulations are validated against Utah FORGE downhole temperature measurements and are benchmarked alongside probabilistic (Gaussian

Process) and direct data-driven (XGBoost) surrogates in a single controlled THM dataset, enabling an objective comparison of physics-guided structure versus uncertainty-aware and black-box (purely data-driven) alternatives.

This advancement provides a practical link between decline analysis and modern data-driven methods. It means users can benefit from fast, scenario-based ML forecasting while still preserving physical meaning in the results. Together, these models offer a flexible toolkit: DCA for detailed case-by-case understanding, GPR for fast screening with uncertainty, and the equation-informed model for physically realistic predictions that support design optimization and sensitivity studies.

## 8. Conclusions

Conventional decline-curve analysis transferred from petroleum engineering often fails to represent geothermal temperature behavior because it omits heat exchange with surrounding rock and the finite equilibrium temperature, while high-fidelity coupled simulators remain expensive for rapid screening across many EGS designs.

This study establishes physics-guided forecasting framework by generalizing the full Arps decline models with an equilibrium-temperature term consistent with Newton-type cooling. This modification enforces finite late-time behavior appropriate for geothermal systems and reduces exactly to the classical Arps equations when the equilibrium temperature is set to zero, preserving backward compatibility.

Across the controlled thermo-hydro-mechanical dataset, the extended decline models reproduce temperature decline with near-perfect fidelity (median $R^2$ = 0.999; median RMSE = 0.071 °C), with the stretched-exponential form selected most often (60%) followed by the hyperbolic form (37%). Field validation against Utah FORGE temperature measurements supports the ability of the modified formulations to capture observed decline while enforcing finite equilibrium behavior.

Building on these physics-based curves, an equation-informed neural network embeds the modified decline equations as internal computational layers to predict decline parameters from five design and operating inputs and render full 0-60 month temperature profiles that retain analytical structure. Typical hold-out errors are MAE = 3.06 °C and RMSE = 4.49 °C, indicating reliable full-trajectory forecasting while preserving interpretability.

A direct multi-horizon Gaussian Process Regression surrogate provides rapid, uncertainty-aware forecasts every 3 months from 3 to 60 months and delivers robust generalization on hold-out cases (macro $R^2$ ≈ 0.965; RMSE ≈ 3.39 °C; MAE ≈ 2.34 °C), with predictive intervals that quantify uncertainty across horizons.

Direct XGBoost models serve as a deterministic, non-physics-informed baseline: general trends are reproduced, but errors are higher (around 7 °C RMSE), reinforcing the value of embedding physical structure or uncertainty modeling for dependable geothermal forecasting.

**Nomenclature**

$B_f$: Thermal expansion coefficient of fluid, 1/°C

$B_s$: Thermal expansion coefficient of solid, 1/°C

B: Skempton pore pressure coefficient

$c_f$: Fluid compressibility, $psi^{-1}$. (1 $psi^{-1}$ = 1.45038×$10^{-4}$ $Pa^{-1}$).

$C_T$: Thermal diffusivity, $m^2/s$

D: Decline rate constant

G: Shear modulus, Pa

k: Reservoir permeability, md

$k_T$: Rock thermal conductivity, $W \cdot m^{-1} \cdot °C^{-1}$

$p_r$: Reservoir pore pressure, Pa

T(t): Temperature as a function of time t, °C

$T_{inj}$: Injection temperature, °C

$T_{res}$: Formation temperature, °C

$T_0$: Initial temperature, °C

$T_\infty$: Asymptotic temperature, °C

$v_u$: Undrained Poisson's ratio

v: Drained Poisson's ratio

$\sigma_x$: Principal stress in x-direction, Pa

$\sigma_{xy}$: Shear stress, Pa

$\sigma_y$: Principal stress in y-direction, Pa

α: Biot's coefficient

μ: Fluid viscosity, cp

ϕ: Reservoir porosity, %